\documentclass[%
 reprint,
 amsmath,amssymb, 
 aps, pre, 
floatfix %, draft
]{revtex4-2}

\usepackage{graphicx}% Include figure files
\usepackage{bm}% bold math
\usepackage[colorlinks=true,linkcolor=blue,citecolor=blue,urlcolor=blue]{hyperref}% add hypertext capabilities

\usepackage{todonotes}
\usepackage{lineno}

\usepackage{placeins}

\newcommand{\supplfigref}[1]{Fig.~S{#1} of the SI}
\newcommand{\new}[1]{ #1}

\newcommand{\beginsupplement}{%
        \setcounter{figure}{0}
        \renewcommand{\thefigure}{S\arabic{figure}}%
     }

\begin{document}

\renewcommand*{\figureautorefname}{Fig.}
\renewcommand*{\sectionautorefname}{Sec.}
\newcommand*{\myeqref}[2][Eq.~]{%
  \hyperref[{#2}]{#1(\ref*{#2})}%
}

\title{On the Quality of Uncertainty Estimates from  Neural Network Potential Ensembles}
\author{Leonid Kahle}
\email{lkahle@materialsdesign.com}
 \altaffiliation[Present address: ]{Materials Design Inc., San Diego, California 92131, USA}
\author{Federico Zipoli}
\email{fzi@zurich.ibm.com}
\affiliation{
National Centre for Computational Design and Discovery of Novel Materials MARVEL, %Cognitive Computing and Computational Sciences Department, 
IBM Research Europe, % S\"aumerstrasse 4,  %CH-8803 R\"uschlikon,
Zurich, Switzerland}
\date{\today}

\begin{abstract}
Neural network potentials (NNPs) combine the computational efficiency of classical interatomic potentials with the high accuracy and flexibility of the \textit{ab initio} methods used to create the training set, but can also result in unphysical predictions when employed outside their training set distribution. Estimating the epistemic uncertainty of a NNP is required in active learning or on-the-fly generation of potentials.
Inspired from their use in other machine-learning applications, NNP ensembles have been used for uncertainty prediction in several studies, with the caveat that ensembles do not provide a rigorous Bayesian estimate of the uncertainty. To test whether NNP ensembles provide accurate uncertainty estimates,  we train such ensembles in four different case studies, and compare the predicted uncertainty with the errors on out-of-distribution validation sets.
Our results indicate that NNP ensembles are often overconfident, underestimating the uncertainty of the model, and require to be calibrated for each system and architecture. We also provide evidence that Bayesian NNPs, obtained by sampling the posterior distribution of the model parameters using  Monte-Carlo techniques, can provide better uncertainty estimates.
\end{abstract}

\maketitle

\section{Introduction}

Interatomic potentials (IPs) are a long-established method to describe the potential energy and forces in atomic systems, and have provided important insights into the physics of atomic structures in the last seven decades~\cite{alder_radial_1955,rahman_correlations_1964,ercolessi_interatomic_1994,gonzalez_force_2011}.
A notable advantage of IPs over first-principles approaches such as density functional theory (DFT)~\cite{hohenberg_inhomogeneous_1964, kohn_self-consistent_1965} and first-principles molecular dynamics~\cite{car_unified_1985} is their higher computational efficiency, allowing the simulation of larger systems over longer time scales.
In recent years, machine-learning techniques have been successfully applied to develop force fields for atomistic simulations~\cite{ceriotti_atomistic_2019, musil_physics-inspired_2021,behler_four_2021}.
Among other machine-learning techniques, (deep) neural networks (NN) can be used as force and energy predictors, resulting in  neural network  potentials (NNPs),
which have led to significant advances in atomistic simulation, combining high computational efficiency with great flexibility and accuracy~\cite{zuo_performance_2020,li_complex_2020,qi_bridging_2021}.
One assumption behind IPs is that the energy of the system can be described as a sum over atomic energies: $E = \sum_i E_i$.
Another assumption behind most NNPs is that the atomic energy is a function of the local neighborhood of that specific atom, that is all atoms within a radial cutoff distance, $r_c$.
Any IP, and therefore also NNP, should be invariant to translation, rotation, and permutation of atoms in the system.
This is most commonly achieved by constructing a descriptor of the atomic environment that displays those symmetries.

One approach to generate NNPs is to use NNs to learn parameters of a physics-inspired functional form, for example to tackle long-range electrostatic  contributions to energy and forces~\cite{ghasemi_interatomic_2015, bleiziffer_machine_2018, nebgen_transferable_2018, deng_electrostatic_2019}, which should be relevant in ionic systems. However, recent work has shown that NNPs are capable of accurate predictions in ionic systems without explicitly including long-range interactions~\cite{marcolongo_simulating_2020, huang_deep_2021}.
NNPs can also be used directly for prediction of energies, forces, and stresses, without enforcing a particular functional form.
As a first example, Behler and Parrinello implemented a NNP~\cite{behler_generalized_2007} using atomic symmetry functions as an SO(3) invariant descriptor.
Another example is \texttt{DeePMD} developed by Zhang \textit{et al.}, where the relevant descriptors are learned during the training and SO(3) invariance is encoded in the mathematical form of the first layers~\cite{zhang_deep_2018}.

Convolutional neural networks and graph convolutional neural networks achieve invariance via the application of tailored convolution filters~\cite{duvenaud_convolutional_2015,hy_predicting_2018,ryczko_convolutional_2018, xie_crystal_2018, klicpera_directional_2020}, one prominent example being Schnet~\cite{schutt_schnet_2017, schutt_schnet_2018}.
A recent addition to the plethora of neural network potentials are tensor-field
NNPs,  which encode SO(3) equivariance in the convolutional operations~\cite{thomas_tensor_2018, mailoa_fast_2019}.
Batzner \textit{et al.} released \texttt{NequIP}~\cite{batzner_se3-equivariant_2021}, an efficient implementation of a NNP using equivariant convolutions.

While the architecture of the network and the type and size of the input descriptor are of great importance to build an accurate NNP, training the model with an extensive data set of high quality is just as relevant.
One common approach is to sample configurations with \textit{ab initio} methods, such as Born-Oppenheimer molecular dynamics at different thermodynamic conditions~\cite{zhang_end--end_2018, zuo_performance_2020}.
Another possibility is an active-learning approach, a repeated cycle of exploration, labeling, and (re)training of a model as follows:
\begin{enumerate}
    \item \textit{exploring} configurations space, e.g., \textit{via} molecular dynamics, using a NNP obtained in the  previous iteration or from an initial training.
    \item \textit{labeling} a subset of configurations using an \textit{ab initio} method.
    \item \textit{training} a new NNP with the labeled configurations.
\end{enumerate}
As a first example, Marcolongo \textit{et al.}~\cite{marcolongo_simulating_2020}
trained a NNP in this iterative fashion to model Li-ion diffusion in solid-state electrolytes, where the configurations in the second step were selected randomly from the molecular-dynamics trajectory of the first step, and were therefore sampled according to the Boltzmann distribution.
A possible disadvantage of this method is that data selected randomly could be redundant, adding no new information to the training set.
Several works in the past have concentrated on detecting configurations that provide additional data that are not already covered  by the training set.~\cite{li_molecular_2015, miwa_interatomic_2017}.
This can either enable active exploration of configuration space to obtain better coverage, or so-called on-the-fly training during a molecular-dynamics or Monte-Carlo trajectory, proposed as early as 1997 by Vita and Car~\cite{vita_novel_1997} and used or developed further in subsequent work~\cite{csanyi_learn_2004}.
\new{Using NNPs in such a scenario can be challenging, as it is very hard to predict when a model is outside the training set distribution~\cite{zhang_active_2019,wen_uncertainty_2020,schran_committee_2020}.
Neural networks are often described as ``universal function approximators''.
That flexibility  directly results in the difficulty to control prediction trends when departing from the training set distribution.
}

Detecting configurations that provide additional data translates to detecting atomic environments of high epistemic uncertainty.
While aleatoric uncertainty is due to noise inherent in the labels, epistemic uncertainty
is due to lack of data and is the subject of this work~\cite{kendall_what_2017, tagasovska_single-model_2019, abdar_review_2021}.
In the remainder, we  simply refer to uncertainty, meaning epistemic uncertainty originating from data scarcity.
One approach to estimate the uncertainty of the prediction of neural networks relies on the prediction of several instances, so-called  neural-network ensembles, where the uncertainty is estimated from the deviation of the output of different models that were trained on the same data.
The members of the ensemble ideally have different architectures or, as a minimal criterion, have the same architecture but are initialized independently.
Ensemble networks led to promising results in many machine-learning applications~\cite{lakshminarayanan_simple_2017, pearce_uncertainty_2020, yao_quality_2019}.
Ensemble NNPs were, for example, used by Zhang \textit{et al.}~\cite{zhang_active_2019}, but also
other work has relied  on similar approaches to select configurations of high model uncertainty~\cite{chen_iterative_2020,miwa_interatomic_2017,jeong_efficient_2020}.
However,  the uncertainty derived from ensembles is not, strictly speaking, a Bayesian uncertainty estimate~\cite{yarin_gal_uncertainty_2016,pearce_uncertainty_2020,dangelo_repulsive_2021}.
The same applies to dropout techniques, which have also been used to estimate uncertainty in NNPs~\cite{wen_uncertainty_2020}.
The method relying on the least approximations to obtain uncertainty estimates are Bayesian NNs~\cite{mackay_practical_1992,neal_bayesian_1996,martin_computing_2020}, obtained by sampling the posterior distribution of NN parameters.
Bayesian NNPs, which do not rely on ensembles or dropout, have not yet been developed to the best of our knowledge, most likely due to the increased complexity and high computational cost of exploring the posterior distribution of the NNP parameters using expensive sampling techniques such as  Hamiltonian Monte Carlo~\cite{neal_bayesian_1993,chandra_langevin-gradient_2018,baldock_bayesian_2019,yao_quality_2019}.

In this work, we explore the relationship between the uncertainty predicted by NNP ensembles and the true error of the prediction, showing
that ensembles are prone to common bias and overconfident predictions, depending on model architecture and the system,
requiring careful calibration of the uncertainty.
Furthermore, we compare choosing configurations based on the predicted uncertainty, as proposed by Zhang \textit{et al.}~\cite{zhang_active_2019}, to random sampling during exploration, as done by Marcolongo \textit{et al.}~\cite{marcolongo_simulating_2020}.
In \autoref{sec:methods} we explain the methods used in this work.
The results are shown and discussed in \autoref{sec:results} and
we present our conclusion in \autoref{sec:conclusions}.

\section{Methods}
\label{sec:methods}

\subsection{Systems and NNPs}

\new{
In order to allow for general conclusions, we used training and validation data from four very different atomic systems:
an atomic dimer, an aluminum surface slab, bulk liquid water, and a benzene molecule in vacuum.
The training and validation data were produced with a Lennard-Jones potential, density-functional theory, or the highly accurate coupled-cluster single-, double-, and perturbative triple-excitations method, CCSD(T).
The neural-network architectures we tested were a custom-built neural network, the \texttt{DeePMD}~\cite{zhang_deep_2018} framework, and the \texttt{NequIP}~\cite{batzner_se3-equivariant_2021} potential.
}

As a first case study, we trained a NNP on the potential energy surface (PES) of an atomic dimer, i.e., two atoms of the same species in vacuum. For this purpose, we chose a parametrized model as the ground truth, namely the Lennard-Jones IP, where the energy is a function of the interatomic distance $r$:
\begin{equation}
    E(r) = 4\epsilon \left[\left( \frac{\sigma}{r}\right)^{12} - 
    \left( \frac{\sigma}{r}\right)^{6} \right],
\end{equation}
where we set $\epsilon=10.34\,\text{meV} = 120\,\text{K} \cdot k_B^{-1} $ and $\sigma=0.34$~nm (which are the  parameters used by Rahman~\cite{rahman_correlations_1964} to study atomic correlation in liquid argon).
In the range of interest (0.3 to 0.7~nm) we chose randomly ten training points according the Boltzmann weight $p(E)\propto e^{\frac{E}{k_B T}}$ with $T=20\,$K.
\new{Such a simple system allows for a significant reduction of the dimensionality of the problem and for a custom-built NNP, leading to better interpretability of the results.}
%For this simple case study %, the atomic dimer, 

We implemented a custom NNP using the \texttt{pytorch} framework, with an input size of one (the interatomic distance), a hidden layer of 64 neurons, and an output layer of one (the energy).
A neural network requires a non-linear activation function, and common choices of such activation functions are 
the hyperbolic tangent \texttt{tanh}, the \texttt{sigmoid} function,  rectified linear units (\texttt{ReLU}),
continuously differentiable exponential linear units (\texttt{CELU})~\cite{barron_continuously_2017}, and Gaussian error linear units (\texttt{GELU}).
These activation functions are shown in \supplfigref{1}.
We implemented NNPs with each type of activation function, as well as one NNP where the activation function of each neuron was randomly chosen among aforementioned functions.
We trained each NNP with the Adam optimizer for 50'000 steps with a learning rate of $10^{-3}$.

We used the smooth edition of \texttt{DeePMD}~\cite{zhang_deep_2018}, as implemented in the \texttt{deepmd-kit} package (version 1.3.3), to build a potential for bulk and surface aluminum (Al) and for liquid water.
The \texttt{DeePMD} NNPs had three layers for the  local-embedding network, their sizes being 32, 64, and 128, respectively, and three layers for the fitting network, each consisting of 256 neurons.
The training was performed using stochastic-gradient descent with an exponentially declining learning rate.

To have a more realistic and complex scenario than the atomic dimer, we trained the \texttt{DeePMD} potential on forces and energies from an Al(100) surface with an adatom of the same species.
\new{The bulk portion of the system is representative of metals, while the presence of the surface leads to additional complexity.}
Extensive FPMD simulations of this system were performed by Nguyen \textit{et al.} of a system of 295 atoms~\cite{nguyen_adatom-induced_2018}, consisting of six layers of Al (49 atoms per layer) with an additional atom on the surface.
The calculations were performed with Born-Oppenheimer molecular dynamics as implemented in the Quantum ESPRESSO (QE) distribution~\cite{giannozzi_quantum_2009} using the Perdew-Burke-Ernzerhof (PBE) exchange-correlation functional~\cite{perdew_generalized_1996} in the canonical (NVT) ensemble.
The temperatures investigated were 300 -- 600 K in steps of 100 K, and additionally 800 and 1000 K, with $\approx$30~ps of dynamics collected for every simulation.
We selected evenly distributed snapshots from each trajectory: 144 snapshots from the trajectory at 1000~K, 72 snapshots at 800~K, and 36~snapshots from the simulations at 300 -- 600~K.
For each temperature, 12 snapshots were retained for validation in order to obtain a temperature-dependant validation error. 
We used the remaining snapshots (training set) to train a NNP using \texttt{DeePMD}~\cite{zhang_deep_2018}, with the parameters given in~\supplfigref{2}.
An additional 91 snapshots were created from an exploration with the NNP at 1000~K and added to the training set.
Training the \texttt{DeePMD} NNP with this data set, we obtained a trained model $\mathcal{M}_{Al}$.
One hundred additional snapshots were sampled by exploring with $\mathcal{M}_{Al}$ the same system at 1500~K in order to obtain an additional validation set, $\mathcal{D}^{Al}_{1500}$.

We also trained a NNP with \texttt{DeePMD} for the commonly studied system of bulk water, \new{representative of highly ergodic liquid systems with high directionality of bonds, resulting in the complex chemistry of water.}
We used the data set created by Cheng \textit{et al.}~\cite{cheng_ab_2018,cheng_ab_2019} of 64 water molecules in a cubic periodic cell.
We split the data set into four equally large data sets $\mathcal{D}^{H_2O}_{0\dots 3}$ of 300 configurations each according to the potential energy of the configuration,
where the potential energy $E_i$ of a configuration in set $k$ is smaller than (or equal to) the energies of  configuration in set $k+1$:
\begin{equation}
 E_i \leq E_j \forall i \in \mathcal{D}^{H_2O}_{k} \land \forall j \in \mathcal{D}^{H_2O}_{k+1}.
\end{equation}
\new{The energy distribution of the four sets of configurations is shown in \supplfigref{8}.}
The model was trained with  $\mathcal{D}^{H_2O}_{0}$ and $\mathcal{D}^{H_2O}_{1\dots 3}$ were kept as validation sets.
The parameters employed for \texttt{DeePMD} (Al and  water) are given in \supplfigref{2}.

Last, we  employed \texttt{NequIP}, developed  by Batzner \textit{et al.}~\cite{batzner_se3-equivariant_2021}, to fit a potential for the benzene molecule in vacuum,
\new{which is a good representative of covalently bonded systems.}
Our \texttt{NequIP} NNP had three convolution layers, and eight was the dimension of the hidden layer and the number of features.
We used a data set of 1500 configurations of C$_6$H$_6$, produced by Chmiela \textit{et al.}~\cite{chmiela_towards_2018}, who
 sampled configurations using MD in the canonical (NVT) ensemble at 500~K, and re-calculated forces and energies using CCSD(T) for this data set. %\footnote{The data set was downloaded at }.
We ordered the set of 1500 configurations by potential energy and split the data into six batches of equal size, obtaining independent sets $\mathcal{D}^{C_6H_6}_{0\dots 5}$ of 250 configurations each.
\new{The energy distribution of the six sets of configurations is shown in \supplfigref{9}.}
Also in this case, the potential energy $E_i$ of a configuration in set $k$ is smaller (or equal) to the energies of  configuration in set $k+1$:
\begin{equation}
 E_i \leq E_j \forall i \in \mathcal{D}^{C_6H_6}_{k} \land \forall j \in \mathcal{D}^{C_6H_6}_{k+1}
\end{equation}
As in the case study of liquid water, we trained the NNP on the set of lowest-energy configurations ($\mathcal{D}^{C_6H_6}_{0}$), and used $\mathcal{D}^{C_6H_6}_{1\dots 5}$ as validation sets. 
The Adam optimizer was used for the training, with a learning rate of 0.01, and the training was stopped after 500 epochs (complete passes over the training data).
All parameters are reported in \supplfigref{3}.

\subsection{NNP ensembles}

Unless stated differently, we trained $M=8$ NNPs with the same data to estimate the ensemble uncertainty, which we found to be converged at that ensemble size.
These models were initialized independently \textit{via} the input of different seeds, which affected both the parameter initialization and the order of sample selection during stochastic gradient descent, where applicable.
From the ensemble of NNPs, we computed the mean prediction and the standard deviation $\sigma$ for every force component $F_{I\alpha}$: 
\begin{align}
    \bar{F}_{I\alpha} &=\frac{1}{M} \sum_m^M F^m_{I\alpha}\\
    \sigma_{I\alpha} &= \sqrt{\frac{1}{M} \sum_m^M \left(F^m_{I\alpha} - \bar{F}_{I\alpha} \right)^2},
\label{eq:deviation-forces}
\end{align}
where the subscripts $I$ and $\alpha$ give atomic indices and spatial dimensions, respectively, and $m$ iterates over models of the ensemble.
$\sigma_{I\alpha}$ is the predicted model uncertainty of the force in direction $\alpha$ of atom $I$.

The error $\epsilon$ of the model was calculated (on a validation/test set) from the difference of the predicted (mean) force $\bar{F}_{I\alpha}$ and the \textit{ab initio} ${F}^{ai}_{I\alpha}$ force (the label):
\begin{equation}
    \epsilon_{I\alpha} = | F^{ai}_{I\alpha} - \bar{F}_{I\alpha}|.
\end{equation}

\subsection{Bayesian NNP}

The simplicity of the NNP for the atomic dimer and the small amount of training data (ten one-dimensional training points) allowed for a Bayesian estimate of the posterior probability of the weights $\theta$, given the data $D=\{(x_i, y_i)\}$:
\begin{equation}
 p(\theta | D) = \frac{p(D| \theta) p(\theta)}{p(D)},
 \label{eq:bayes}
\end{equation}
where $p(D| \theta)$ is the likelihood of the data, $p(\theta)$ is the prior, and $p(D)$ the marginal likelihood.
The likelihood of the data given the parameters $\theta$ of a function $f$, assuming Gaussian white noise of variance $\sigma^2$ in the observations, is given by
\begin{align}
\notag p(D| \theta) =& \prod_{i=1}^n \frac{1}{\sqrt{2\pi\sigma^2}} e^{-\frac{(y_i-f(x_i|\theta))^2}{2\sigma^2}} \\
\notag =& \left(\frac{1}{\sqrt{2\pi\sigma^2}}\right)^n e^{-\frac{\sum_{i=1}^n \left(y_i - f\left(x_i|\theta\right) \right)^2}{2\sigma^2} }\\
%=& \left(\frac{1}{\sqrt{2\pi\sigma^2}}\right)^n e^{-\frac{n}{2\sigma^2} \mathcal{L}(D)} \\
\propto & e^{-\frac{\mathcal{L}(D|\theta)}{T} },
\label{eq:likelihood}
\end{align}
where $\mathcal{L}(D|\theta) = \frac{1}{n}\sum_i^n(y_i-f(x_i|\theta))^2$ is the mean-squared error loss function and $n$ the number of samples.
 
The prior of the parameters, $p(\theta)$, was set to the normal distribution $\mathcal{N}(0, \sigma_p^2)$ with $\sigma_p^2=1000$, which we evaluated \textit{via} a curvature criterion.
We ignored $p(D)$, the marginal likelihood, since it is independent of $\theta$.
Inserting Eq.~\eqref{eq:likelihood} and the Gaussian prior into Eq.~\eqref{eq:bayes}, 
we obtain an energy or cost function of the parameters of the network 
that can be sampled at a fictitious temperature $T$~\cite{mackay_practical_1992,chandra_langevin-gradient_2018,baldock_bayesian_2019,wenzel_how_2020}:
\begin{equation}
p(\theta|D,T) \propto e^{-\frac{1}{T} \mathcal{L}(D|\theta)  - \frac{\theta^2}{2\sigma_p^2}}.
\end{equation}
We sampled this landscape using Hamiltonian Monte Carlo (HMC)~\cite{neal_mcmc_2011}, with the Hamiltonian dynamics being integrated with the velocity Verlet integrator~\cite{verlet_computer_1967}.
We set the fictitious temperature to 0.05 (resulting in a cold posterior~\cite{wenzel_how_2020}) and used 10'000 steps of Hamiltonian dynamics, with a timestep of 0.015, which resulted in an acceptance rate of 0.675 of Monte-Carlo moves, close to the optimal acceptance rate of 0.65~\cite{neal_mcmc_2011}.
We sampled the parameter distribution by selecting 100 models from the MC trajectory to calculate mean and uncertainty as for the NNP ensembles.

\subsection{Statistical analysis}

We interpret the ensemble uncertainty $\sigma_{I\alpha}$ 
as a predictor for the error
$\epsilon_{I\alpha}$.
A straightforward approach, also employed by Zhang \textit{et al.}~\cite{zhang_dp-gen_2020},
is to classify configurations based on an uncertainty threshold~$\sigma_{max}$.
Configurations with all $\sigma_{I\alpha}< \sigma_{max}$ were classified as low error- and were classified as high-error configurations otherwise.
The condition was given by  $\epsilon_{I\alpha}< \epsilon_{max}$.
We note that $\sigma_{max}$ and $\epsilon_{max}$ did not have to be equal, in order to account for systematic underestimates.
We formulated two distinct requirements for the uncertainty:
\begin{enumerate}
    \item A requirement for accurate results is that configurations that produce a high error $\epsilon_{I\alpha} > \epsilon_{max}$, due to lack of training data in that region of configuration space are detected \textit{via} a high uncertainty, $\sigma_{I\alpha} > \sigma_{max}$.
    \item A computational requirement is to achieve high precision for high-error configurations, in order to avoid falsely selecting many configurations that are described well by the model.
\end{enumerate}

The first requirement is given by the true positive rate (TPR), also called recall or sensitivity, which is given by the 
ratio of correctly positive prediction among all elements with the following condition:
\begin{equation}
    TPR = \frac{TP}{TP+FN},
    \label{eq:tpr}
\end{equation}
where TP and FN refer to true positives ($\epsilon_{I\alpha}  > \epsilon_{max}$ and $\sigma_{I\alpha}  > \sigma_{max}$) and false negatives ($\epsilon_{I\alpha} > \epsilon_{max}$ and $\sigma_{I\alpha} < \sigma_{max}$), respectively. 
The second requirement, stating that we want to maximize the ratio of true high-error configuration  to high-uncertainty configurations, is  described by another metric, the precision or positive predictive value (PPV):
\begin{equation}
    PPV = \frac{TP}{TP+FP},
    \label{eq:ppv}
\end{equation}
where $FP$ refers to the false positives ($\epsilon_{I\alpha} < \epsilon_{max}$ and $\sigma_{I\alpha}  > \sigma_{max}$).

\section{Results and discussion}
\label{sec:results}

\subsection{The atomic dimer}
\begin{figure}[t]
    \centering
    \includegraphics[width=\hsize]{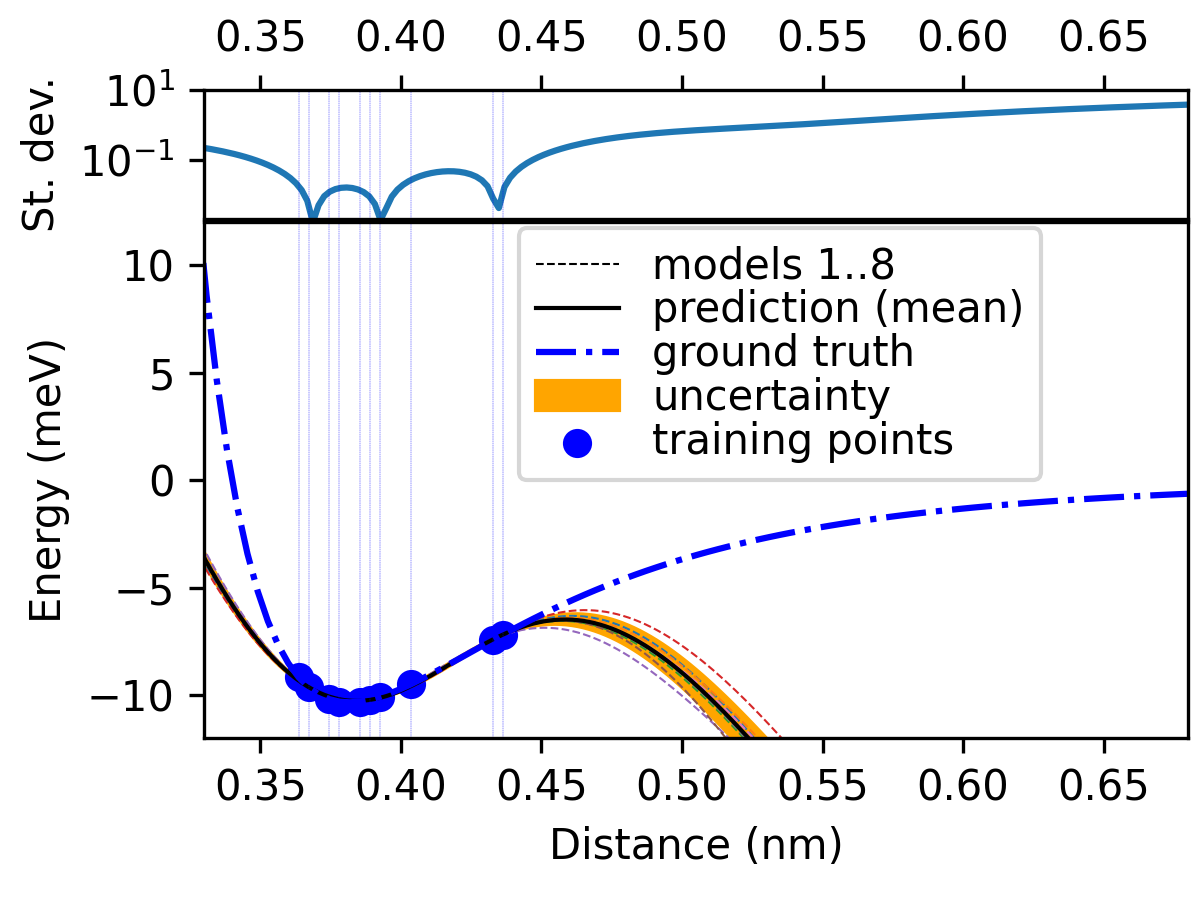}
    \caption{The ground truth (blue dash-dotted lines) and the training points (blue dots and dotted vertical lines) are shown together with predictions from eight models trained on the same data but initialized differently. 
    In the top panel we plot the logarithm of the standard deviation between the models.
    The model uses the hyperbolic tangent as an activation function.
    The top panel shows the standard deviation of the model prediction as a function of $r$ on a semi-logarithmic scale.}
    \label{fig:lj-Tanh}
\end{figure}

\begin{figure}[t]
    \centering
    \includegraphics[width=\hsize]{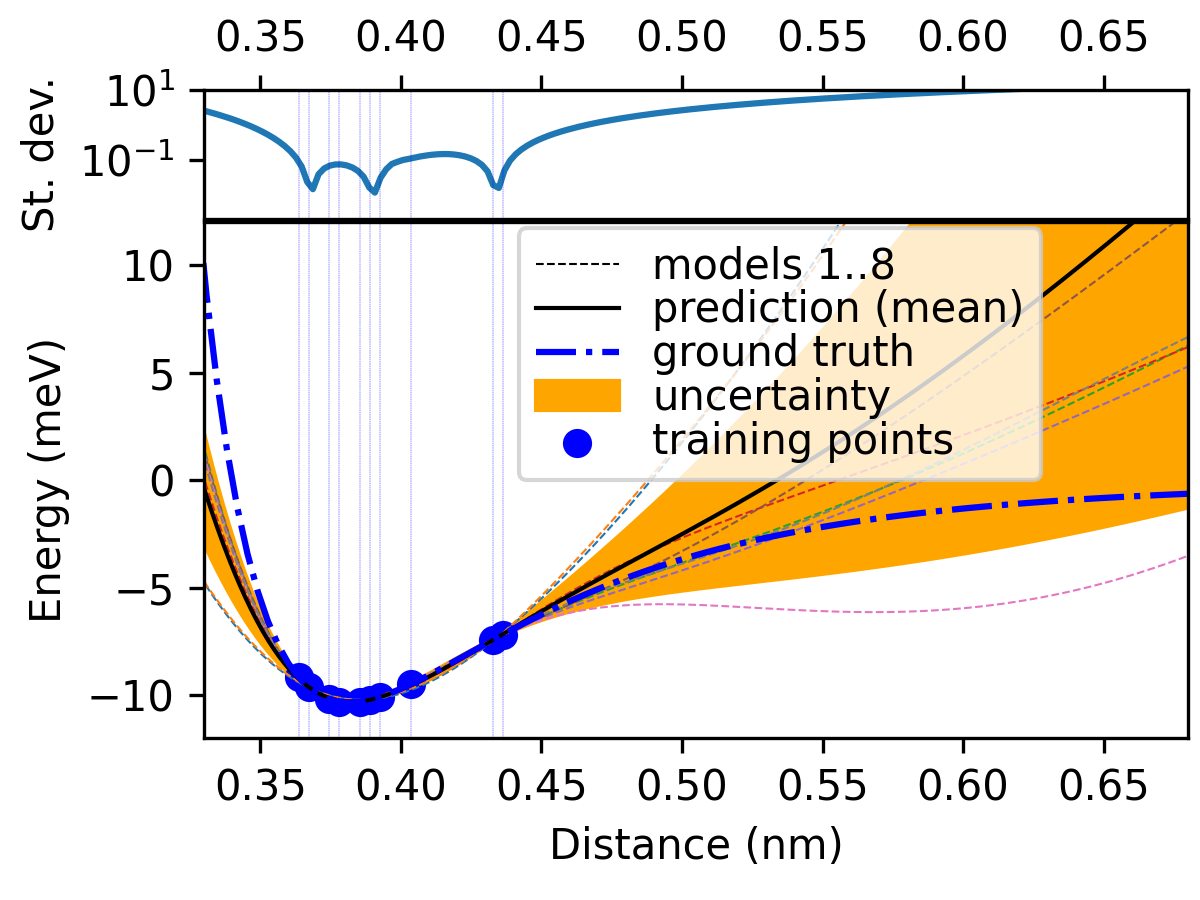}
    \caption{Similar to \autoref{fig:lj-Tanh}, training data (blue dots) sampled from the ground truth (blue dash-dotted line) were used to train different models (dotted lines). 
    The models are initialized independently and also have different architectures, ensured \textit{via} a random assignment of activation function to every neuron in the hidden layer.
    The estimate of the uncertainty is improved due to reduced bias.}
    \label{fig:lj-rand}
\end{figure}

\begin{figure}[b]
    \centering
    \includegraphics[width=\hsize]{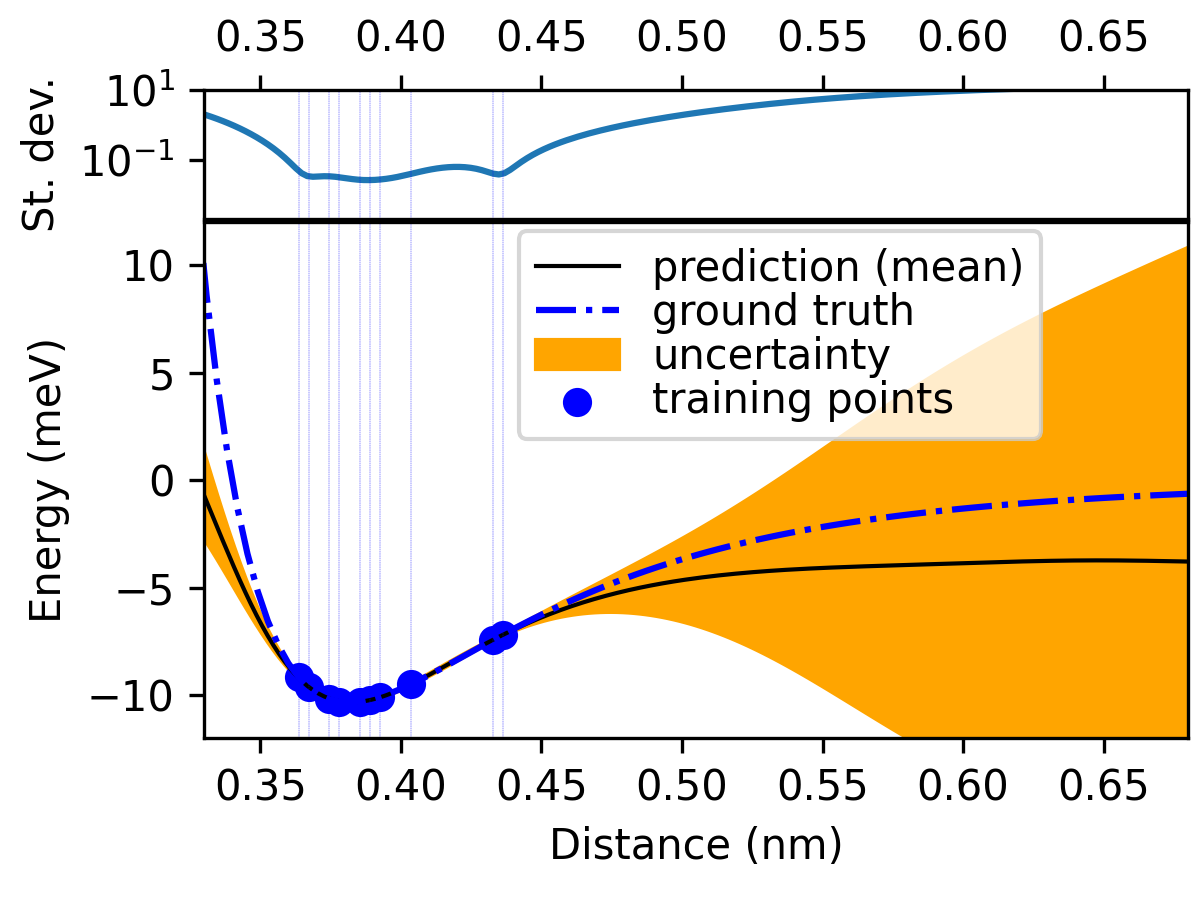}
    \caption{Similar to \autoref{fig:lj-Tanh}, the training data (blue dots) are sampled from the ground truth (blue dash-dotted line).
    The dotted lines show the predictions of models sampled from the posterior parameter distribution, and the orange area depicts one standard deviation from the mean.
    }
    \label{fig:lj-bayesian}
\end{figure}

The custom NNP for the atomic dimer receives a scalar input (the distance), and predicts the energy, and has 64 hidden units.
The predictions of this model using the hyperbolic tangent (\texttt{tanh})  as the non-linear activation function are shown for the  range of $r$ from 0.33 to 0.68~nm in \autoref{fig:lj-Tanh}.
The NNP ensemble mean is very accurate at predicting the PES in the region of training data, which we call the interpolation regime.
In the extrapolation regime, where $r<0.35$~nm or $r>0.45$~nm, no data are supplied and the models deviate from the ground truth (blue dash-dotted lines).
However, the members of the ensemble deviate in a very similar fashion, most probably due to common biases.
This is not only the case when using this particular activation function: all activation functions (\texttt{ReLU}, \texttt{CELU}~\cite{barron_continuously_2017}, \texttt{GELU}, and \texttt{sigmoid})  show the same behavior (see \supplfigref{4}).
However, different activation functions result in different biases in the region of larger distances ($r>0.45$~nm).
Using \texttt{tanh} or \texttt{sigmoid} as activation functions leads to underestimates of the energy at $r>0.45\,$nm, whereas other activation functions (\texttt{ReLU}, \texttt{CELU}) lead to overestimates. 
We suspect that this behavior is because the networks with the latter activation functions are prone to extrapolate the slope from the closest training points.
\new{Using the uncertainty of the ensemble as a predictor for low accuracy~\cite{zhang_active_2019} would result in this case study in false positives in data-scarce regions due to the bias in the model.
%In such a scenario, with low predictiveness of the ensemble uncertainty, sampling randomly configurations for a new training set is the easier approach.
}

The  NNP ensemble where every neuron within the hidden layer is randomly assigned one out of the  different activation functions we study (\texttt{tanh}, \texttt{sigmoid}, \texttt{GELU}, \texttt{ReLU}, and \texttt{CELU}) produces
predictions that are more diverse (see \autoref{fig:lj-rand}) outside the training set distribution.
The ensemble displays higher variance of the output in the extrapolative region of $r>0.45$~nm, which we interpret as evidence that  bias has been reduced.
The region of smaller distances ($r<0.35$~nm) is  biased towards slopes of smaller magnitude. 
Compared to the results of the ensembles with the same architecture, diversifying the NNP ensemble results in lower bias, which is also discussed by Jeong \textit{et al.}~\cite{jeong_efficient_2020}.
\new{
The uncertainty of ensembles with a varying model architecture is a better predictor for the accuracy of the model when predicting unseen data and could be useful in active learning.
Current approaches employing NNP ensembles~\cite{zhang_active_2019,chen_iterative_2020} do not use varying architectures of the ensemble members to reduce the collective bias.
The results of this simple case study of the atomic dimer indicate that this could help in improving the uncertainty estimate.
}

As a last example, we estimated the posterior distribution of the weights using HMC.
The Bayesian model predicts (see 
\autoref{fig:lj-bayesian}) a high uncertainty in the extrapolation regime, for both $r<0.35$ and $r>0.45$, and also an increased uncertainty in the region around $r=0.42$~nm due to lower data density.
The Bayesian model compares well to ensembles with the same and with varying architecture, as the bias is further reduced.
The uncertainty predicted by the Bayesian NNP resembles most closely our expectation due to the sharp increase of the uncertainty when extrapolating, and a moderate increase in uncertainty when interpolating in regions of low training data density.

\subsection{Al(100) surface}
\label{sec:al}

\begin{figure}[b]
    \centering
    \includegraphics[width=\hsize]{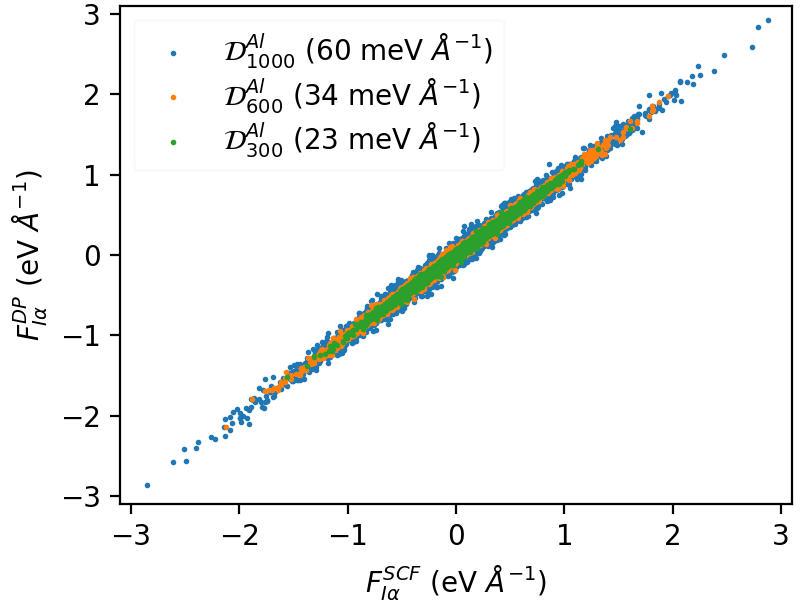}
    \caption{The prediction of a force component ($y$ axis) against the DFT-calculated force ($x$ axis) for the validations sets $\mathcal{D}^{Al}_{300}$, $\mathcal{D}^{Al}_{600}$, and $\mathcal{D}^{Al}_{1000}$ in blue, orange, and green respectively.
    The RMSE of the forces is given in brackets in the legend.}
    \label{fig:al-force-force}
\end{figure}

\begin{figure}[t]
    \centering
    \includegraphics[width=\hsize]{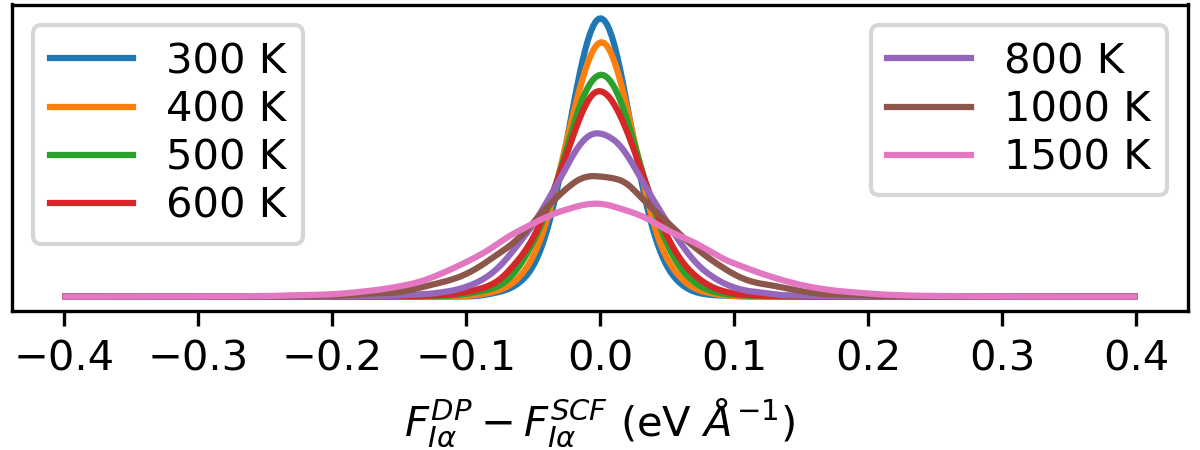}
    \caption{Histograms of the difference of the predicted and calculated force components reveal an approximately Gaussian-distributed error with increasing deviation with higher temperature.}
    \label{fig:al-errors}
\end{figure}

\begin{figure}[b]
    \centering
    \includegraphics[width=\hsize]{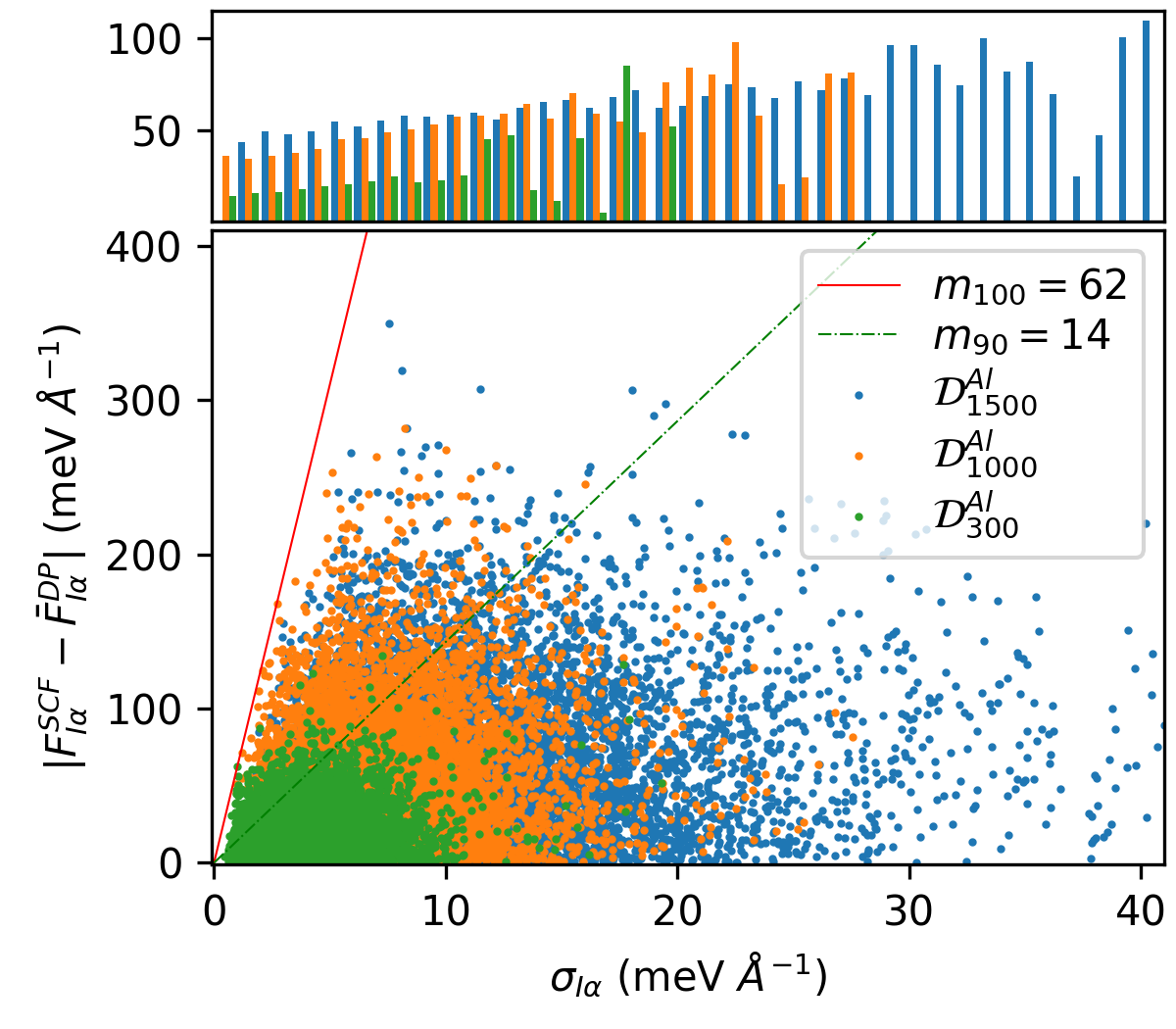}
    \caption{
    The ensemble uncertainty $\sigma_{I\alpha}$ (calculated via \myeqref{eq:deviation-forces}) is plotted against the  error $\epsilon_{I\alpha}$, calculated as the squared difference between the predicted value and the label, for three validation set $\mathcal{D}^{Al}_{1500}$, $\mathcal{D}^{Al}_{1500}$, and $\mathcal{D}^{Al}_{1500}$ in blue, orange, and green, respectively.
    The red solid line gives the line of smallest slope the bounds the data points from above.
    In the top panel, the reliability diagram shows the mean error for all data points with a given uncertainty, with the same color encoding.
    }
    \label{fig:al-error-prediction}
\end{figure}

\begin{figure}[t]
    \centering
    \includegraphics[width=\hsize]{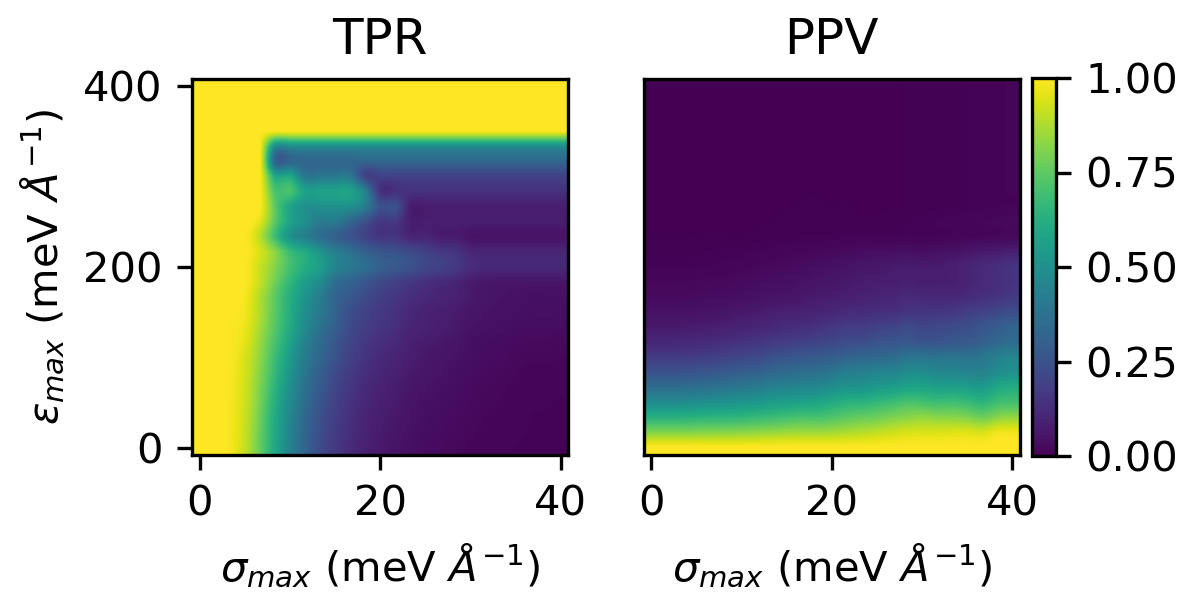}
    \caption{For the validation set $\mathcal{D}^{Al}_{1500}$ we plot the recall or TPR as a heat map over different true errors and predicted errors (left).
    On the right, we plot the precision or PPV for the same data set.}
    \label{fig:tpr-ppv-al}
\end{figure}

\begin{figure}[t]
    \centering
    \includegraphics[width=\hsize]{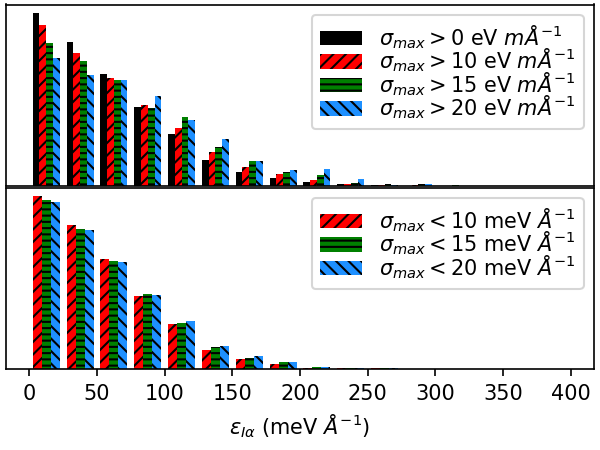}
    \caption{The true error distribution of snapshots is plotted for all configurations of the  validation set  $\mathcal{D}^{Al}_{1500}$ for configurations that have a predicted uncertain above (top panel) or below (bottom panel) a given threshold. The black bars display the true error histogram for all configurations. The histograms being very similar to each other shows that for $\mathcal{D}^{Al}_{1500}$ it is not possible to single out high-error configurations via a threshold on the model uncertainty.}
    \label{fig:hist-errors-al}
\end{figure}

\new{The ensemble of eight \texttt{DeePMD} models, trained on energies and forces of atoms in and on an Al surface, results in an accurate NNP within and close to the training set distribution.}
The forces on Al atoms for the configurations of three validation sets ($\mathcal{D}^{Al}_{300}$,  $\mathcal{D}^{Al}_{600}$, and $\mathcal{D}^{Al}_{1000}$) are plotted in \autoref{fig:al-force-force}, showing excellent agreement, evidence that the ensemble is able to capture the interatomic interactions of Al in the bulk and on the surface.
We observe that the validation sets sampled at higher temperatures lead to higher forces and higher root mean square errors (RMSE) in the prediction, 
which is expected because the system explores a larger region of phase space at higher temperatures, which translates to a more complex training problem.
A histogram of the error for each temperature we studied is shown in \autoref{fig:al-errors}, confirming this behavior.
The validation set sampled at 1500~K has the highest error.
We remind the reader that no configurations sampled at that temperature were used in training, and we mark these as out-of-distribution (\textit{ood}) samples.

One question of interest is how to detect these \textit{ood} samples, or -- more generally --  high-error configurations.
The relationship between predicted uncertainty and true error is shown for three validation sets ($\mathcal{D}^{Al}_{1500}$, $\mathcal{D}^{Al}_{1000}$, and $\mathcal{D}^{Al}_{300}$) in \autoref{fig:al-error-prediction}.
We observe that  the  error $\epsilon_{I\alpha}$ is generally about 1 order of magnitude higher than the predicted uncertainty $\sigma_{I\alpha}$.

No perfect correlation between  $\epsilon_{I\alpha}$ and $\sigma_{I\alpha}$  is expected,
and such a correlation is not a necessary criterion to select snapshots with unacceptable error.
A more pragmatic criterion is that a positive constant $c$ can be found such that $\epsilon_{I\alpha} <  c \cdot \sigma_{I\alpha} $ for all (or a significant fraction of) $I\alpha$. 
Such a constant or slope, for example, is visible in Fig.~2(e) of Ref.~\cite{vandermause_--fly_2020}, where the error estimate is given by a Gaussian process regression model.
We draw such a line as a guide to the eye in \autoref{fig:al-error-prediction}, choosing the line of smallest slope above the data points (and going through the origin).
The slope of this line is 62, which means that in order to achieve a very high degree of certainty that the error is below a given threshold $\epsilon_{max}$, the  uncertainty threshold $\sigma_{max}$ has to be almost 2 orders of magnitude above $\epsilon_{max}$.
\new{We also draw the line that lies above 90\% of the  data points, resulting in a slope of 14.
This shows that even if one is willing to accept a significant amount of high-error configurations, the uncertainty threshold nevertheless lies an order of magnitude above the error threshold.
}
A reliability diagram is shown in the top panel of \autoref{fig:al-error-prediction}, where we bin the data points by their uncertainty and plot the mean error for every bin.
The mean error increases when the uncertainty is higher, but there are strong signs of miscalibration.
A well-calibrated model would result in a reliability diagram close to an identity function.
%However, we observe high errors for data points where the ensemble is confident about the prediction.

Similar to previous work~\cite{schran_committee_2020, zhang_dp-gen_2020}, we assume that a maximum true error, $\epsilon_{max}$, exists above which the properties sampled by the dynamics are no longer trustworthy, requiring a need to recalculate this point during active exploration or marking it for labeling.
We classify force components of the validation set $\mathcal{D}^{Al}_{1500}$ set according to the true error $\epsilon_{I\alpha} > \epsilon_{max}$ and try to predict this class using $\sigma_{I\alpha} > \sigma_{max}$.
The TPR, introduced in \myeqref{eq:tpr}, is shown in the left panel of \autoref{fig:tpr-ppv-al}.
For low $\sigma_{max}$ (below 10~meV\,$\AA^{-1}$), it is possible to guarantee that the true error is bounded by a wide range of $\epsilon_{max}$.
In general, the region of high recall or TPR is at the top left of the heat map, towards high $\epsilon_{max}$ and low $\sigma_{max}$.
The region of high computational efficiency is given by a high precision or PPV and is concentrated in the bottom right, towards high $\sigma_{max}$ and low $\epsilon_{max}$.
The only region of high precision or PPV and high recall or TPR is in the bottom left,
however, this is the region where $\epsilon_{max}$ and $\sigma_{max}$ are so low that almost all data points are true positives (see \supplfigref{5}).
In summary, only by giving very strict criteria on the predicted uncertainty is it guaranteed that the true error is bound, which implies that we can only exclude a negligible amount of the data set as low-error points.
In an on-the-fly training scenario, where the low error of force components is important, the number of useful calculations will be very low due to the high number of false positives and the low number of true positives (see \supplfigref{5}).

In an active-learning scenario, finding all high-error points is not necessary as long as a sufficient number of  high-error configurations are discovered to enrich the training set with $ood$ data, allowing us to relax a very strict threshold $\sigma_{max}$.
In such a scenario, it is important to find training points for the next iteration that are more likely to be outside the previous training set than if sampled randomly.
In \autoref{fig:hist-errors-al} we plot histograms of the true error in the force components $\epsilon_{I\alpha}$ when screening for all predictions above $\sigma_{max}$
($\sigma_{max}=0$ therefore includes all force components).
We observe a shift of the histogram towards larger true errors when selecting for components with higher uncertainties; however, this shift is marginal.
This confirms our interpretation of the right panel of \autoref{fig:tpr-ppv-al}, that it is not possible to achieve a high PPV except when declaring almost all data points as high-error points \textit{via} a very low threshold $\epsilon_{max}$.

\new{To summarize, we note that ensemble uncertainties are about 1 order of magnitude lower than the errors.
This observation needs to be accounted for when relying on ensemble uncertainties.
When specifying the error threshold $\epsilon_{max}$, one needs to investigate which uncertainty criterion $\sigma_{max}$ this requires.
As this case study shows, it is possible that $\sigma_{max}$ needs to be significantly larger than $\epsilon_{max}$.
Furthermore, the results of this case study indicate that sampling based on an uncertainty criterion~\cite{zhang_active_2019} or sampling randomly~\cite{marcolongo_simulating_2020} should result in very similar distributions.
Random sampling is, for obvious reasons, easier to implement and computationally more efficient and would therefore be preferable.
}

\subsection{Water}
\label{sec:liq-water}
\begin{figure}[t]
    \centering
    \includegraphics[width=\hsize]{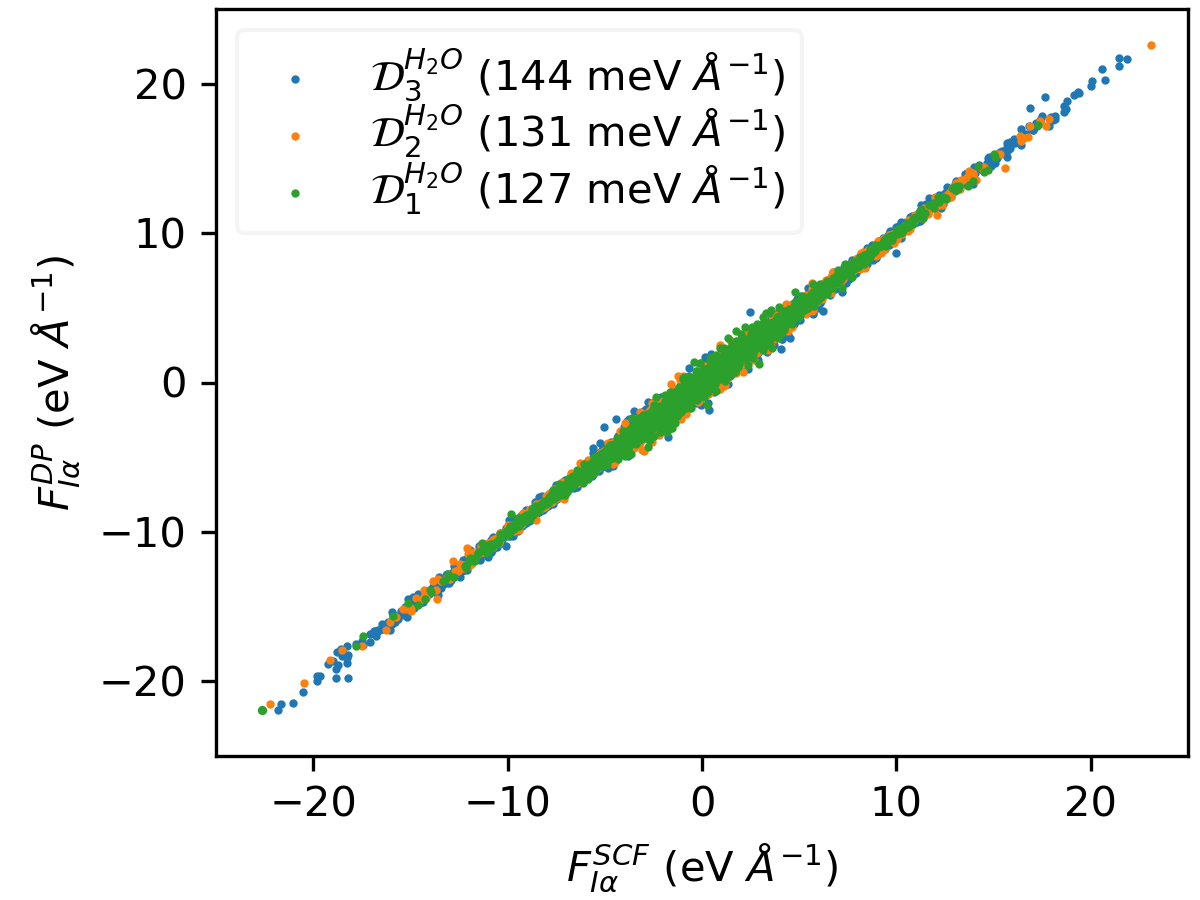}
    \caption{
    We plot the DFT forces against the predicted forces of DeePMD for the validation sets $\mathcal{D}^{H_2O}_{1}$, $\mathcal{D}^{H_2O}_{2}$, and $\mathcal{D}^{H_2O}_{3}$. RMSE is given in brackets inside the legend.}
    \label{fig:force-force-water}
\end{figure}

\begin{figure}[b]
    \centering
    \includegraphics[width=\hsize]{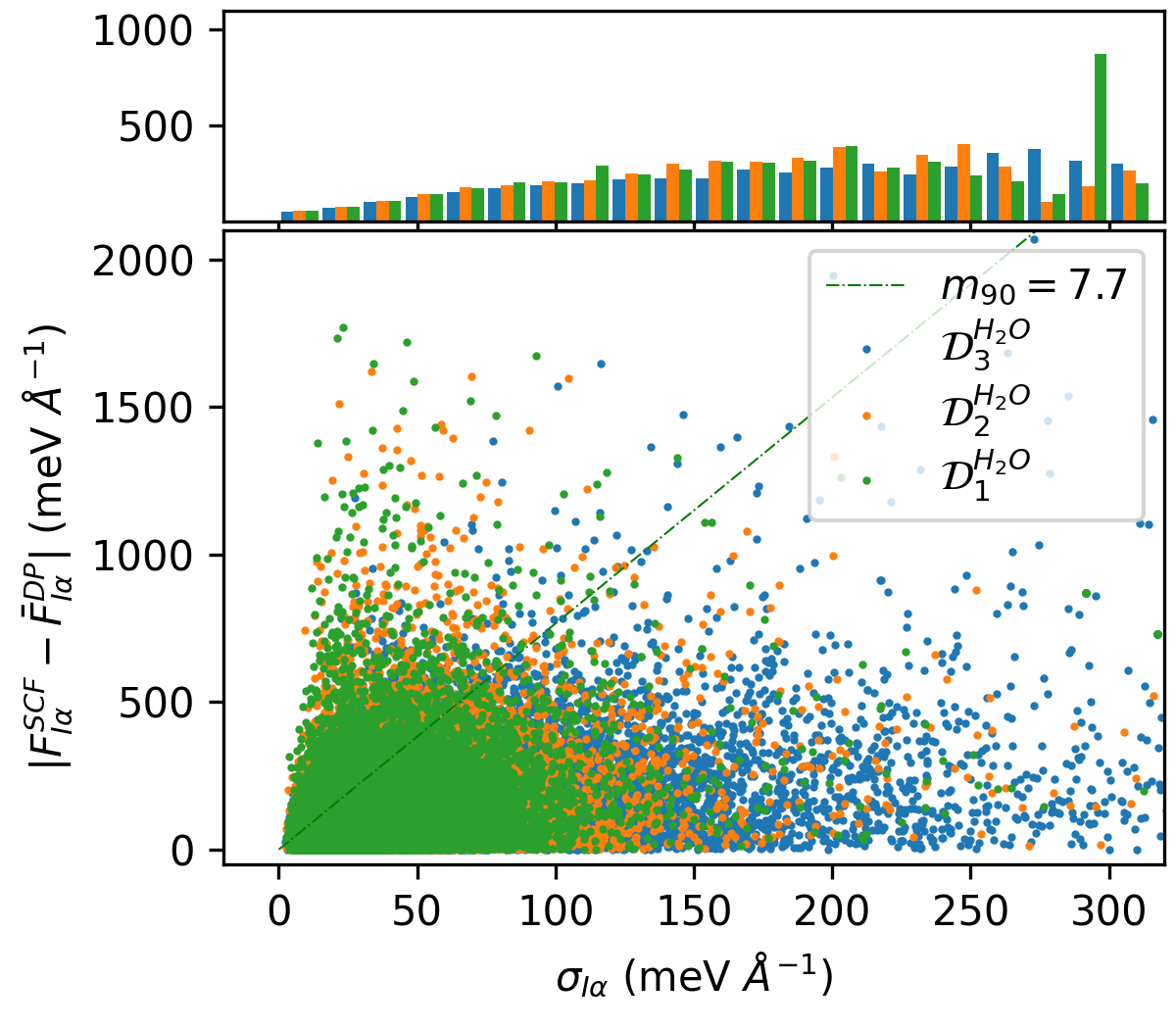}
    \caption{The predicted uncertainty $\sigma_{I\alpha}$ against the true error in liquid water for the validation sets 
    $\mathcal{D}^{H_2O}_{1}$, $\mathcal{D}^{H_2O}_{2}$, and $\mathcal{D}^{H_2O}_{3}$) shown as green, orange, and blue dots, respectively. %
    }
    \label{fig:error-prediction-water}
\end{figure}

\begin{figure}[t]
    \centering
    \includegraphics[width=\hsize]{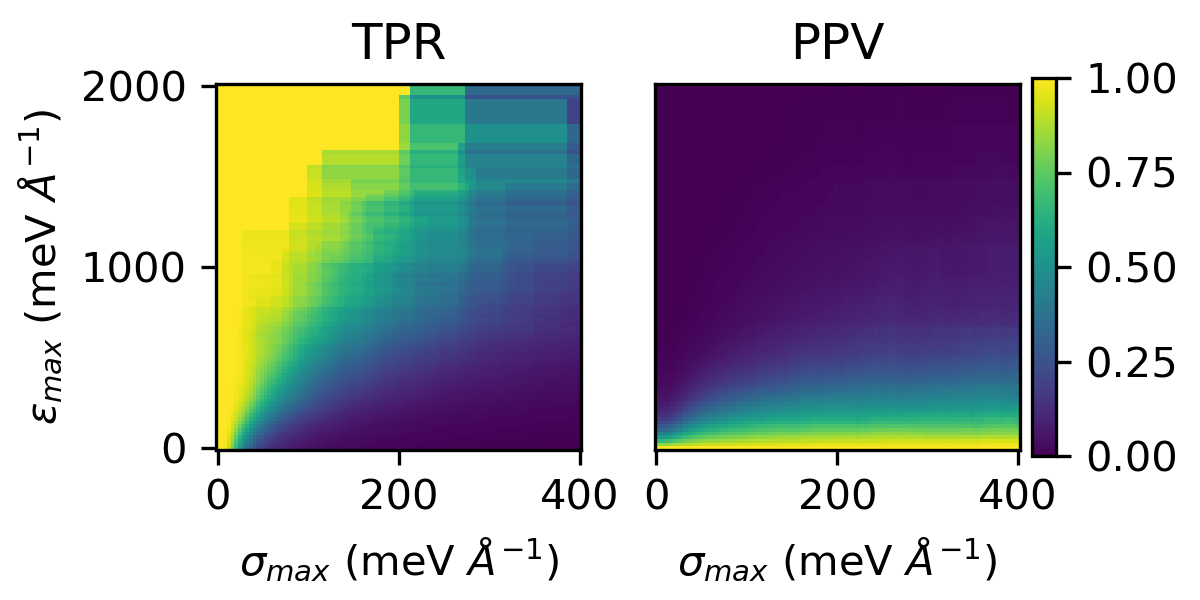}
    \caption{Recall (TPR) and precision (PPV)  for the $\mathcal{D}^{H_2O}_{3}$ validation set for liquid water.}
    \label{fig:tpr-ppv-water}
\end{figure}

\begin{figure}[b]
    \centering
    \includegraphics[width=\hsize]{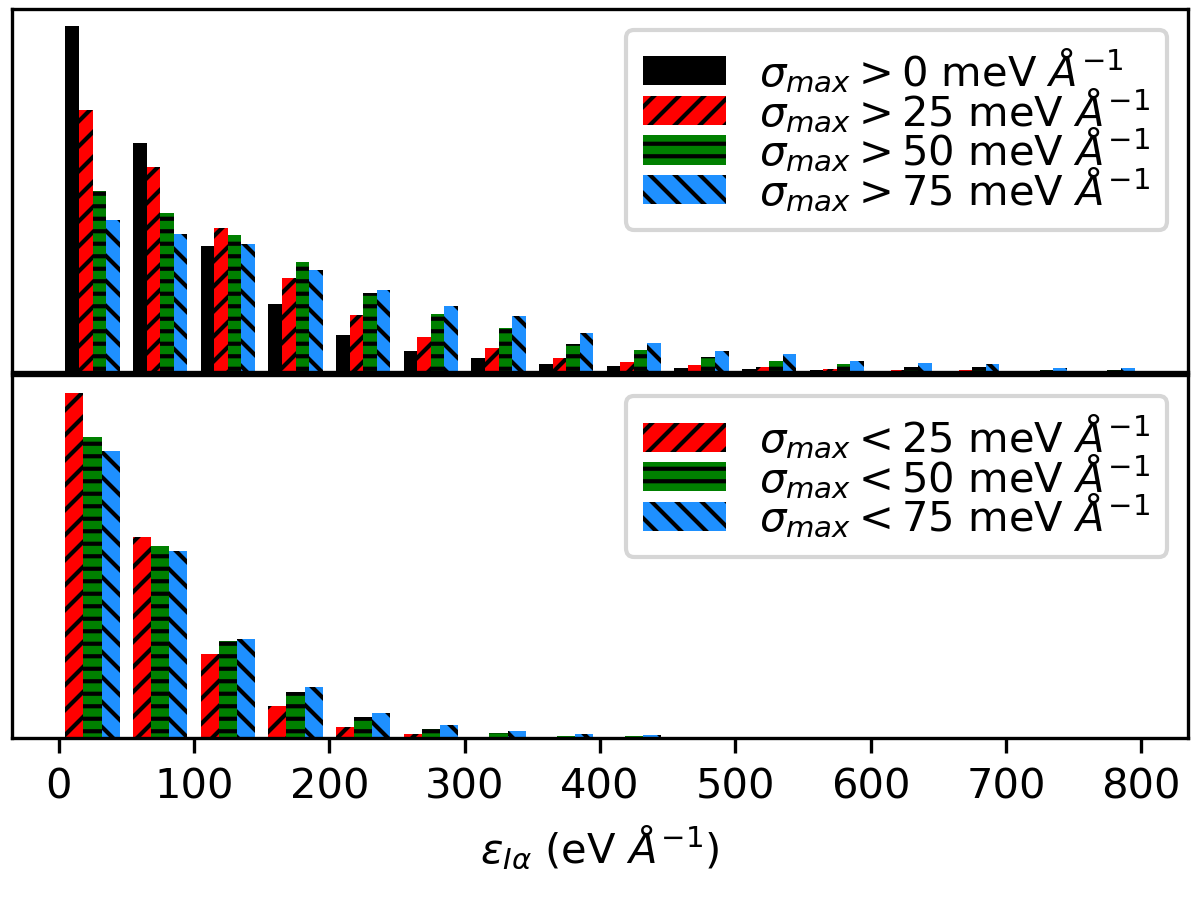}
    \caption{Histogram of the true error for different subsets of $\mathcal{D}^{H_2O}_{3}$, based on the predicted model uncertainty. The black bars show the distribution of error distribution of the entire validation set. With increasing model uncertainty threshold, the histogram becomes  skewed towards higher-error configurations.}
   \label{fig:hist-errors-water}
\end{figure}

In \autoref{fig:force-force-water} we show the forces predicted by the \texttt{DeePMD} NNP ensemble against the labels for the three validation sets of bulk water.
We note that the training and validation configurations do not originate solely from a molecular-dynamics trajectory, but also from different sampling strategies that lead to significantly higher forces~\cite{cheng_ab_2019} than in the previous case study.
The error in the forces increases slightly for snapshots of higher potential energies, and the highest RMSE is found for $\mathcal{D}^{H_2O}_3$.
Nevertheless, the good reproduction of forces in unseen data gives evidence of a very accurate NNP \new{within and close to the training set distribution, gently reducing in accuracy as one moves away from the training examples.}

The predicted uncertainty against the true error for all validation sets, plotted in \autoref{fig:error-prediction-water}, displays a  behavior similar to that in the case study of Al (cf. \autoref{fig:al-error-prediction}).
The first similarity is that the uncertainty predicted by the NNP ensemble is underestimated with respect to the error by the mean prediction, also by approximately 1 order of magnitude.
A distinction, however, is that there is no line of finite slope that bounds all data points from above (i.e., it is not possible to bound the error rigorously),
due to the presence of configurations that have been predicted with zero uncertainty but display finite true errors.
\new{
A line that lies above 90\% of the validation points has a slope of 7.7, which is a bit lower than in the previous case study. Nevertheless, the conclusion is the same as for Al(100): the uncertainty threshold needs to be an order of magnitude higher than the error one is willing to accept.}
Another similarity is the presence of false positives, i.e., force components are predicted relatively accurately despite the high uncertainty associated with that prediction.
The reliability diagram in the top panel of \autoref{fig:error-prediction-water} bears closer resemblance to an identity than in the case of Al (cf.  \autoref{fig:al-error-prediction}).
We speculate that this is because the validation set $\mathcal{D}^{H_2O}_3$ of liquid water is described worse by the ensemble than the validation set of Al(100) ($\mathcal{D}^{Al}_{1500}$).
The fact that the more accurate a model is, the more likely the model is to be overconfident, has also been observed in other work and is believed to constitute a general trend~\cite{guo_calibration_2017}.

As is the case for Al(100), also in liquid water we find regions of high TPR for high errors threshold $\epsilon_{max}$ and low uncertainty bounds $\sigma_{max}$, meaning that a strict uncertainty bound $\sigma_{max}$ is almost guaranteed to find all examples of high true error $\epsilon_{max}$ (see \autoref{fig:tpr-ppv-water}).
However, regions of high TPR are also regions of low PPV, implying that the computational efficiency in that regime is  low.
As a result, there is no region with a  F$_1$ score close to 1 (see \supplfigref{6}), except for the regime of very strict bounds on uncertainty and error.

In \autoref{fig:hist-errors-water}, we plot histograms of the true error distributions for subsets of the validation sets with a minimal (top panel) or maximal (bottom panel) uncertainty.
Comparing to the case study of Al (cf. \autoref{fig:hist-errors-al}), the histograms change more significantly with increased threshold, meaning that one is likelier to select high-error configurations by using an uncertainty criterion (compared to random selection), which confirms the findings above.
The histogram is still peaked at low true errors, but the difference in the histograms gives evidence that the uncertainty criterion can improve the selection of configurations for labeling and retraining in an active-learning scenario.

\new{
In this case study, sampling based on the ensemble uncertainty is shown to be advantageous, compared to random sampling, as the probability of including high-error configurations is increased.
This has to be weighted against the higher computational cost and complexity of the former approach.
Our results for water do not indicate that the true error can be bounded by a limit on the ensemble uncertainty, merely that the probability of picking a high-error configuration is higher if selecting configurations with high ensemble uncertainty, compared to random choice.
As in the previous case study, to rely on the ensemble uncertainty would require calibration, as the predicted model uncertainties are of a magnitude significantly lower than that of the model errors.
}

\subsection{Benzene}

\begin{figure}[t]
    \centering
    \includegraphics[width=\hsize]{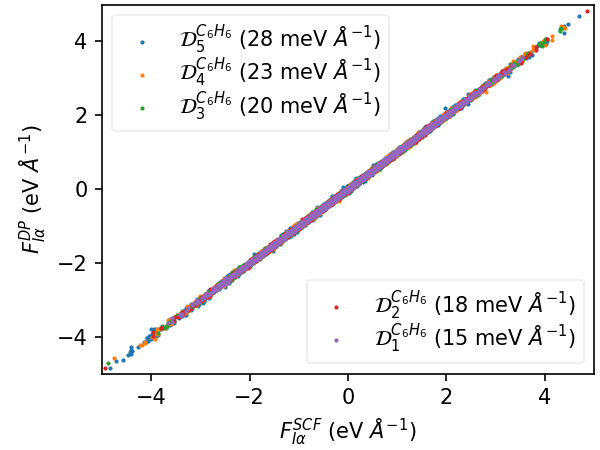}
    \caption{We plot the CCSD(T) forces against the predicted forces of the NNP ensemble for the validation sets $\mathcal{D}^{C_6H_6}_{1\dots 5}$. RMSE is given in brackets inside the legend.}
    \label{fig:force-force-benzene}
\end{figure}

\begin{figure}[b]
    \centering
    \includegraphics[width=\hsize]{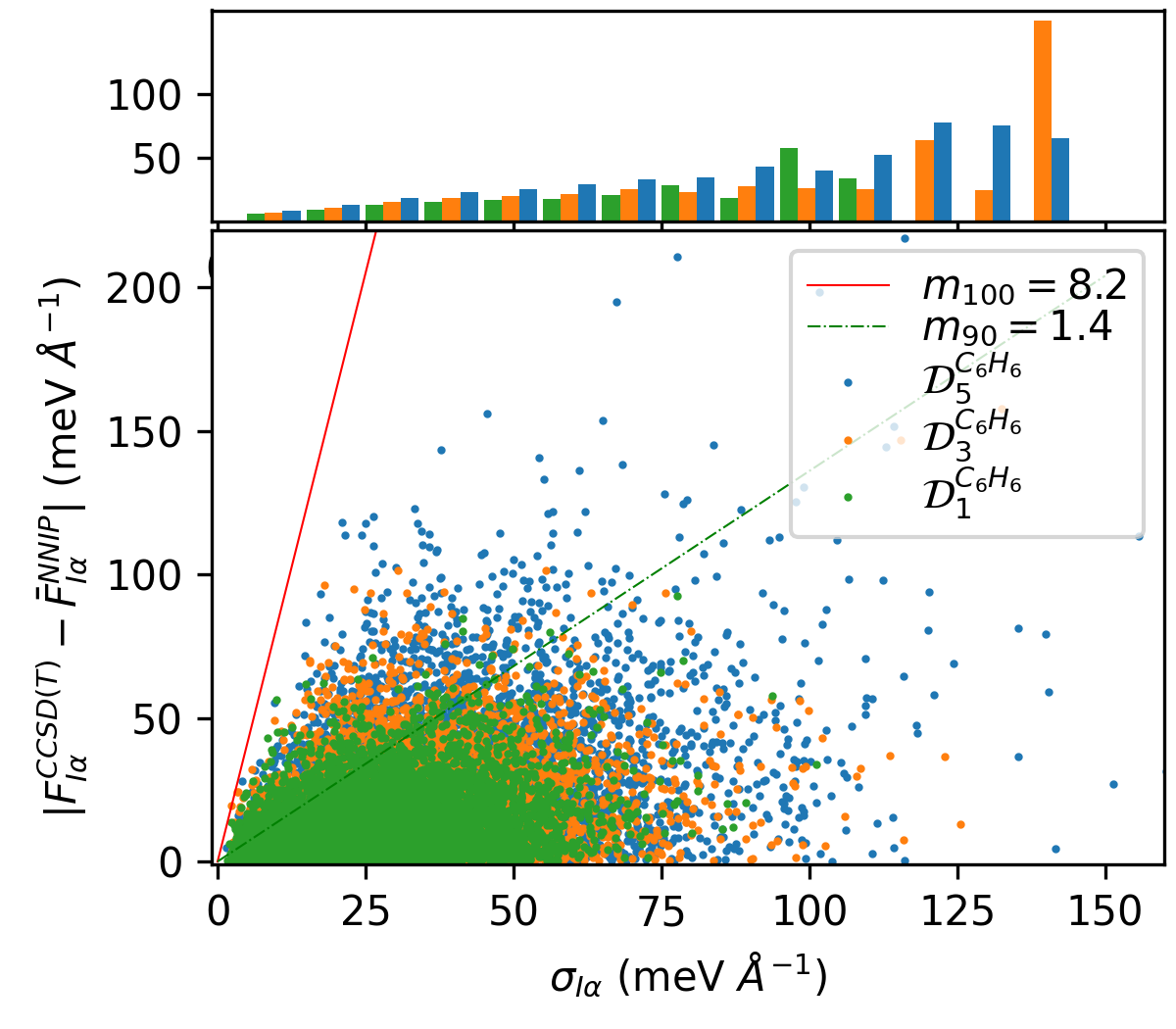}
    \caption{The predicted uncertainty $\sigma_{I\alpha}$ against the true error of an isolated benzene for the validation sets 
    $\mathcal{D}^{C_6H_6}_{1}$, $\mathcal{D}^{C_6H_6}_{3}$, and $\mathcal{D}^{C_6H_6}_{5}$ shown as green, orange, and blue dots, respectively. 
    }
    \label{fig:error-prediction-benzene}
\end{figure}

\begin{figure}[b]
    \centering
    \includegraphics[width=\hsize]{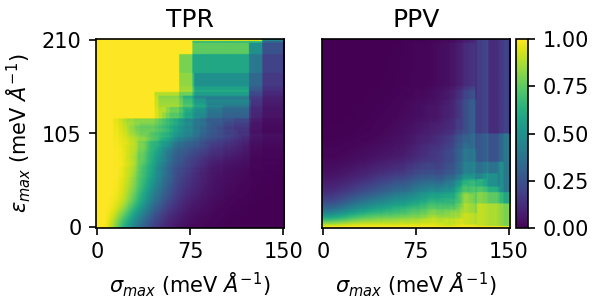}
    \caption{Recall (TPR) and precision (PPV)
    for the $\mathcal{D}_5^{C_6H_6}$ validation set.}
    \label{fig:tpr-ppv-benzene}
\end{figure}

\begin{figure}[t]
    \centering
    \includegraphics[width=\hsize]{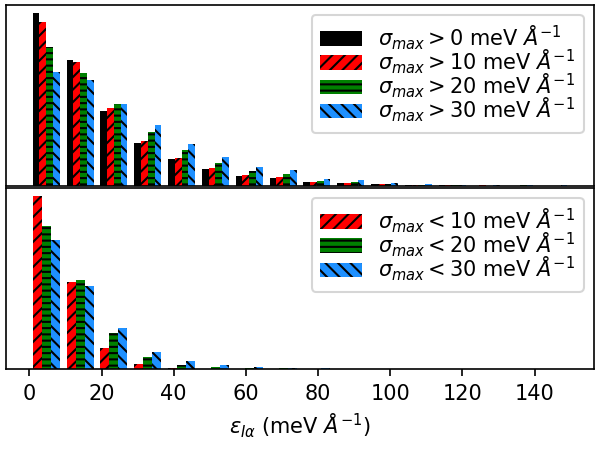}
    \caption{Histogram of the prediction error for different subsets of $\mathcal{D}_5^{C_6H_6}$, selected  based on the ensemble model uncertainty.
    The black bars show the distribution of error distribution of the entire validation set.
    With increasing the model uncertainty threshold, the histogram becomes skewed towards higher-error configurations.
    }
    \label{fig:hist-errors-benzene}
\end{figure}

As a last case study we show the results of the \texttt{NequIP} NNP ensemble trained on the benzene data set.
\texttt{NequIP} generalizes well for this training set and can be trained with a comparatively small number of training points.
The forces estimated by the NNP ensemble trained on the 250 configurations lowest in potential energy are plotted against the  CCSD(T) predictions in \autoref{fig:force-force-benzene} for all validation sets.
The RMSE increases for validation sets of higher potential energy, as in the previous example (cf. \autoref{sec:liq-water}).
\new{The lower complexity of the benzene molecule, compared to water or Al, has allowed us to study which configurations result in comparatively larger model errors.
Based on a simple descriptor of the atomic environment, we find that atoms in the training set whose descriptor falls outside the convex hull of training-set descriptors can have higher errors (see \supplfigref{10}).
}

The uncertainty estimates given by the ensemble deviation are plotted against the error in  \autoref{fig:error-prediction-benzene}.
Here, as for the Al case study (see \autoref{sec:al}), it is possible to heuristically bound the error by a line of slope 8.2, which is significantly smaller than a slope of 62 in the case of Al(100), evidence for a lower degree of overconfidence of the ensemble.
\new{The line that lies above 90\% of the validation points has a slope of 1.4, close to unity.}
Therefore, unlike the previous examples, the true error and predicted uncertainties are of the same order of magnitude, which means that this system and ensemble are better calibrated.
Furthermore, the reliability diagram in the top panel of \autoref{fig:error-prediction-benzene} resembles  the identity function more closely than in the case of Al(100) (cf. \autoref{fig:al-error-prediction}), especially if we exclude the right portion of the histogram where the data points are significantly less.
We observe that, unlike for liquid water and Al(100), the \texttt{NequIP} ensemble is not  overconfident: predictions of an uncertainty $\sigma_{I\alpha}$ have a mean error  $\epsilon_{I\alpha}$ that is of the same order.
Nevertheless, \autoref{fig:tpr-ppv-benzene} implies that the same problem persists as for the uncertainty estimates in Al(100) and liquid water, namely that it is not possible to separate efficiently configurations (or force components) of high uncertainty without also including a large amount of false negatives (configurations that are accurately described, but where the ensemble estimates a high uncertainty).
In \autoref{fig:hist-errors-benzene}, we show the histograms of the error distribution for subsets of the validation set $\mathcal{D}_5^{C_6H_6}$, screened by the ensemble uncertainty.
As for the case of liquid water, the  histogram is shifted towards higher-error configurations when excluding low-uncertainty configurations from the validation set. Nevertheless, the histogram remains with its maximum value at low-error configurations, which means that many false positives are included, and the histograms are not significantly different from a random selection.
\new{
Also in this case study, there is evidence to suggest that sampling based on the uncertainty~\cite{zhang_active_2019} can result in an improved training set with a higher likelihood of unseen new data, compared to random sampling.
%But even in this case, which so far provides the strongest evidence for sampling based on ensemble uncertainty, it is not evident that 
}

To conclude, the \texttt{NequIP} ensemble for benzene is very accurate without being overconfident of the prediction, which is not the case for \texttt{DeePMD} in Al(100) or water.
The focus of this work, however, is not to compare different implementations or architectures of NNPs, which would be a promising question for future research, but rather to highlight that, for certain implementations and systems, NNP ensembles can be overconfident, and that there is high variance in the degree of overconfidence.
The proper calibration of NNP ensembles in order to be accurate without being overconfident needs to be studied further.
A very interesting perspective on this question is provided by Guo \textit{et al.}~\cite{guo_calibration_2017}, who observed that the more accurate a NN model is (due to added parameters and more flexibility), the more it is overconfident.

\section{Conclusions}
\label{sec:conclusions}

Zhang \textit{et al.}~\cite{zhang_dp-gen_2020} have shown substantial evidence that selecting data to label based on the uncertainty of NNP ensemble can result in high-quality models and trajectories.
On the other hand, Marcolongo \textit{et al.} performed an iterative training where samples for the re-training were chosen randomly, i.e., Boltzmann distributed, from a molecular-dynamics trajectory, also resulting in models of high predictive accuracy.
The results of this work reconcile these findings: selecting configurations based on the ensemble uncertainty is in many cases not significantly different from random sampling.
By applying  low thresholds on the uncertainty, the  error of the model can be bound, but this will incur many false positives, resulting in additional computational effort in a learn-on-the-fly or active-learning scenario.

A second finding is that the uncertainty can be significantly underestimated, in the case study of the \texttt{DeePMD} ensemble trained on Al(100)  by an order of magnitude, where
an uncertainty threshold of, e.g., 10~meV\,\AA$^{-1}$ results in errors of $\approx$100~meV\,\AA$^{-1}$.
When employing NNP ensembles, we advise the reader to calibrate the uncertainty threshold on a validation set, as is also advisable in other machine-learning applications~\cite{guo_calibration_2017,seedat_towards_2019}.
Further research is needed on how to properly calibrate NNP ensembles, and an especially interesting avenue is, in our opinion, to explore whether there is an accuracy--confidence trade-off, as observed in other machine-learning applications~\cite{guo_calibration_2017}.

Last, our results indicate that the uncertainty estimates can be improved. 
The NNP ensemble trained on the atomic dimer reveals that ensembles can be biased in a similar manner when the same NN architecture is used, which is also found in similar computational experiments~\cite{yao_quality_2019}.
One method to avoid such common bias is to ensure different architectures of the NNP, for example by randomizing the activation function for every neuron.
A second, more rigorous method, is the implementation of a Bayesian NNP, which can be obtained by sampling the posterior distribution of parameters with HMC and resulted in a significantly improved estimate of the uncertainty.
Training a Bayesian NNP however comes at great complexity and large computational expense, and we defer its implementation and validation to future work.

\section*{Acknowledgements}
This research was supported by the NCCR MARVEL, funded by the Swiss National Science Foundation. 
We are indebted to Simon Batzner and Alby Musaelian for their generous help with \texttt{NequIP}.
L.K. thanks Maria R. Cervera for fruitful discussions.

\bibliography{bibliography}

\onecolumngrid
\newpage
\beginsupplement

\begin{center}
    \large{\textbf{\centering Assessing the Quality of Uncertainty Estimates from Ensembles of Neural Network Interatomic Potentials -- Supplementary Information}}
\end{center}

\vspace{2cm}

\begin{figure}[ht]
    \centering
    \includegraphics[width=\hsize]{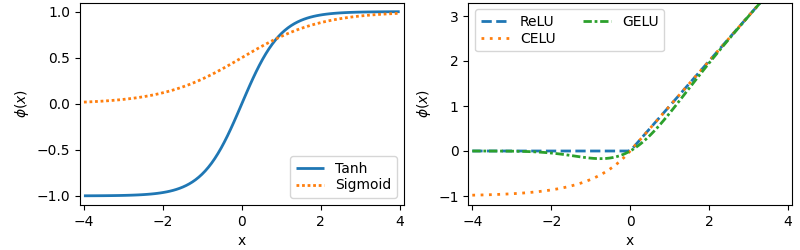}
    \caption{We plot the \texttt{Tanh} and \texttt{Sigmoid} functions in the left panel, both functions that have an asymptotic plateau for $x \rightarrow \pm \infty$.
    In the right panel, we display the activation functions \texttt{ReLU}, \texttt{CELU} and \texttt{GELU}, which have plateaus for $x \rightarrow - \infty$, but have a linear regime for values of $x>>0$ ($x>0$ for \texttt{ReLU}).}
    \label{fig:activation-functions}
\end{figure}

\begin{figure}[ht]
    \centering
    \begin{minipage}[c]{0.49\hsize}
        \begin{verbatim}
{
    "model": {
        "descriptor": {
            "type": "se_a",
            "sel": [300],
            "rcut_smth": RCUT,
            "rcut": RCUT_SMTH,
            "neuron": [32, 64, 128],
            "axis_neuron": 16,
            "resnet_dt": false,
            "seed": SEED0,
        },
        "fitting_net": {
            "neuron": [256,256,256],
            "resnet_dt": true,
            "seed": SEED1,
        }
    },
    "learning_rate": {
        "type": "exp",
        "decay_steps": 5000,
        "decay_rate": 0.95,
        "start_lr": 0.001
    },
\end{verbatim}

    \end{minipage}
     \begin{minipage}[c]{0.49\hsize}
        \begin{verbatim}
    "loss": {
        "start_pref_e": 0.001,
        "limit_pref_e": 1,
        "start_pref_f": 1000,
        "limit_pref_f": 1,
        "start_pref_v": 0.0,
        "limit_pref_v": 0.0
    },
    "training": {
        "systems": ["./data_set"],
        "set_prefix": "set",
        "stop_batch": 200000,
        "batch_size": 1,
        "seed": SEED2,
        "disp_file": "curve.out",
        "disp_freq": 1000,
        "numb_test": 84,
        "save_freq": STOP_BATCH,
        "save_ckpt": "model.ckpt",
        "disp_training": true,
        "time_training": true,
        "profiling": false
    }
}
\end{verbatim}

    \end{minipage}
    \caption{The input for \texttt{DeePMD} used to train NNP ensembles for Al and liquid water.
\texttt{RCUT}/\texttt{RCUT\_SMTH} were set to 8.0/7.0 and 6.5/5.5 for Al and water, respectively.
The different seeds were set by replacing \texttt{SEED0/1/2/3}.}
    \label{fig:supp-deepmd-details}
\end{figure}

\begin{figure}[ht]
    \centering
    \begin{minipage}[c]{0.49\hsize}
        \begin{verbatim}
{
    "seed": SEED,
    "restart": False,
    "append": False,
    "compile_model": True,
    "num_basis": 8, 
    "r_max": 3.5,
    "irreps_edge_sh": "0e + 1o",
    "conv_to_output_hidden_irreps_out": "8x0e", 
    "feature_irreps_hidden": "8x0o+8x0e+8x1o+8x1e",
    "BesselBasis_trainable": True, 
    "nonlinearity_type": "gate", 
    "num_layers": 3,
    "resnet": False,
    "PolynomialCutoff_p": 6,
    "n_train": 100,
    "n_val": 5,
    "learning_rate": 0.01,
    "batch_size": 1,
    "max_epochs": 500,
    "train_val_split": "random",
    "shuffle": True, 
\end{verbatim}
\end{minipage}
    \begin{minipage}[c]{0.49\hsize}
        \begin{verbatim}
    "loss_coeffs": {
        "forces": 0.999,
        "total_energy": 0.001
    }, 
    "atomic_weight_on": False,
    "optimizer_name": "Adam",
    "optimzer_params": {
        "amsgrad": True, 
        "betas": (0.9, 0.999),
        "eps": 1e-08, 
        "weight_decay": 0
    }, 
    "lr_scheduler_name": "CosineAnnealingWarmRestarts", 
    "lr_scheduler_params": {
        "T_0": 10000, 
        "T_mult": 2,
        "eta_min": 0, 
        "last_epoch": -1
    },
    "force_append": False, 
    "allowed_species": tensor([1, 6])
}
\end{verbatim}
\end{minipage}
\caption{The input for \texttt{NequIP} used to train an NNP ensemble for benzene.
The different seeds were set by replacing \texttt{SEED}.}
\label{fig:suppl-nequip-details}
\end{figure}

\begin{figure}[th]
    \includegraphics[width=0.49\hsize]{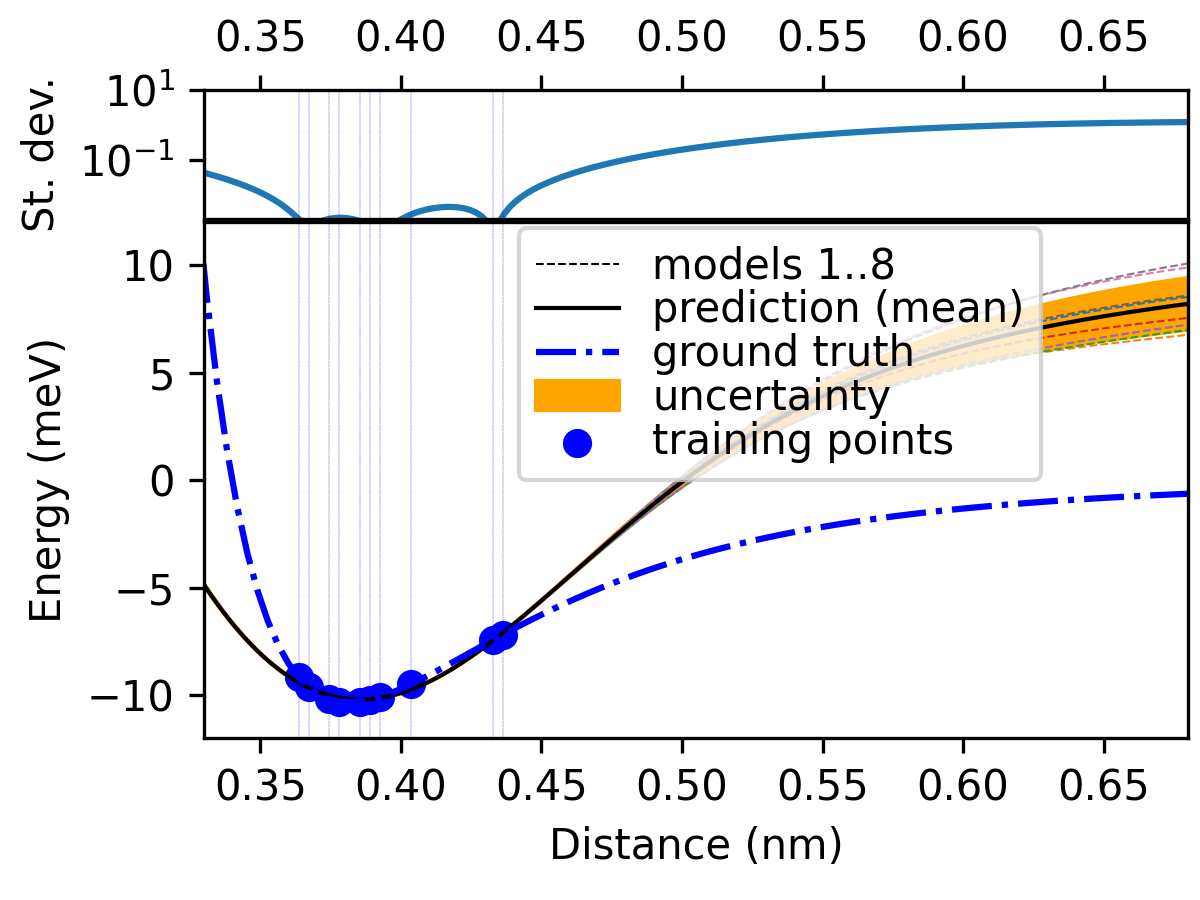}
    \includegraphics[width=0.49\hsize]{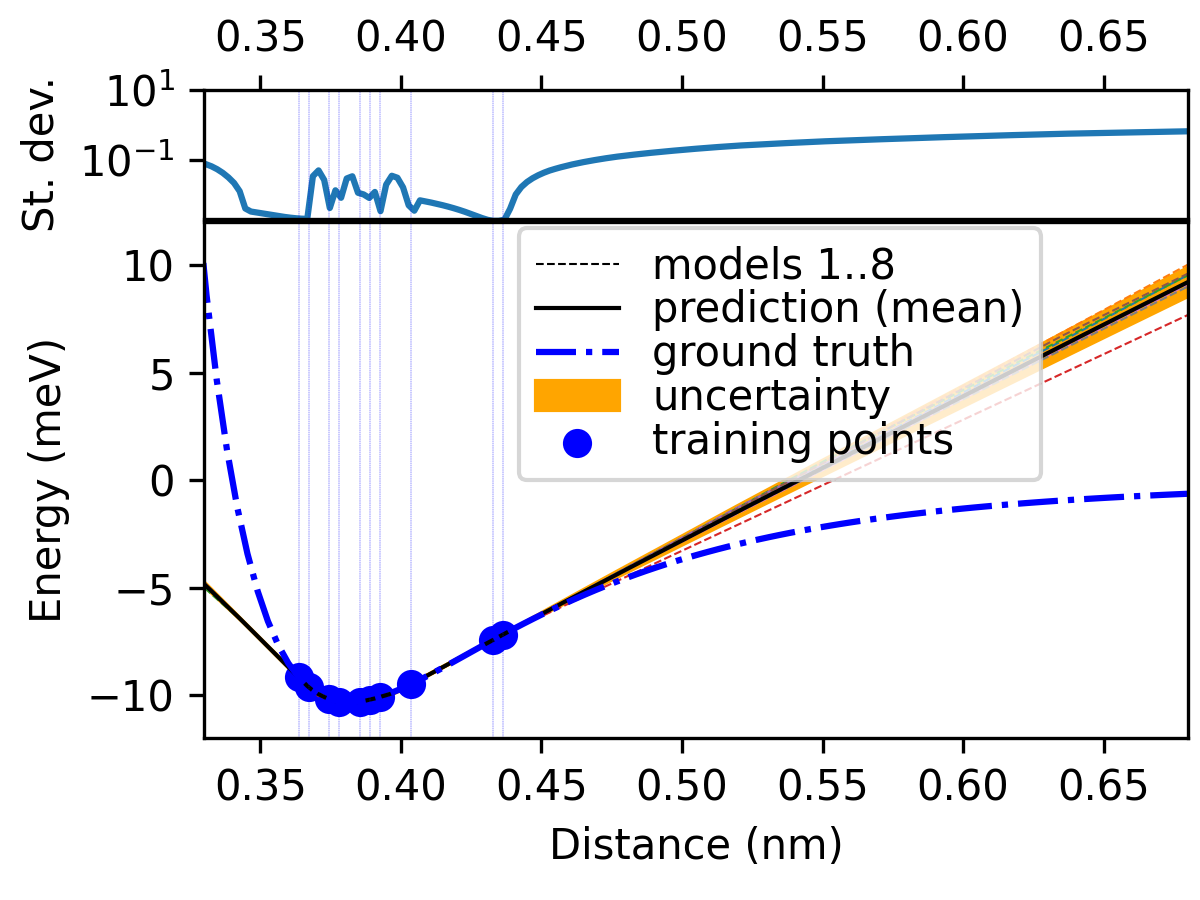}
    \\
    \includegraphics[width=0.49\hsize]{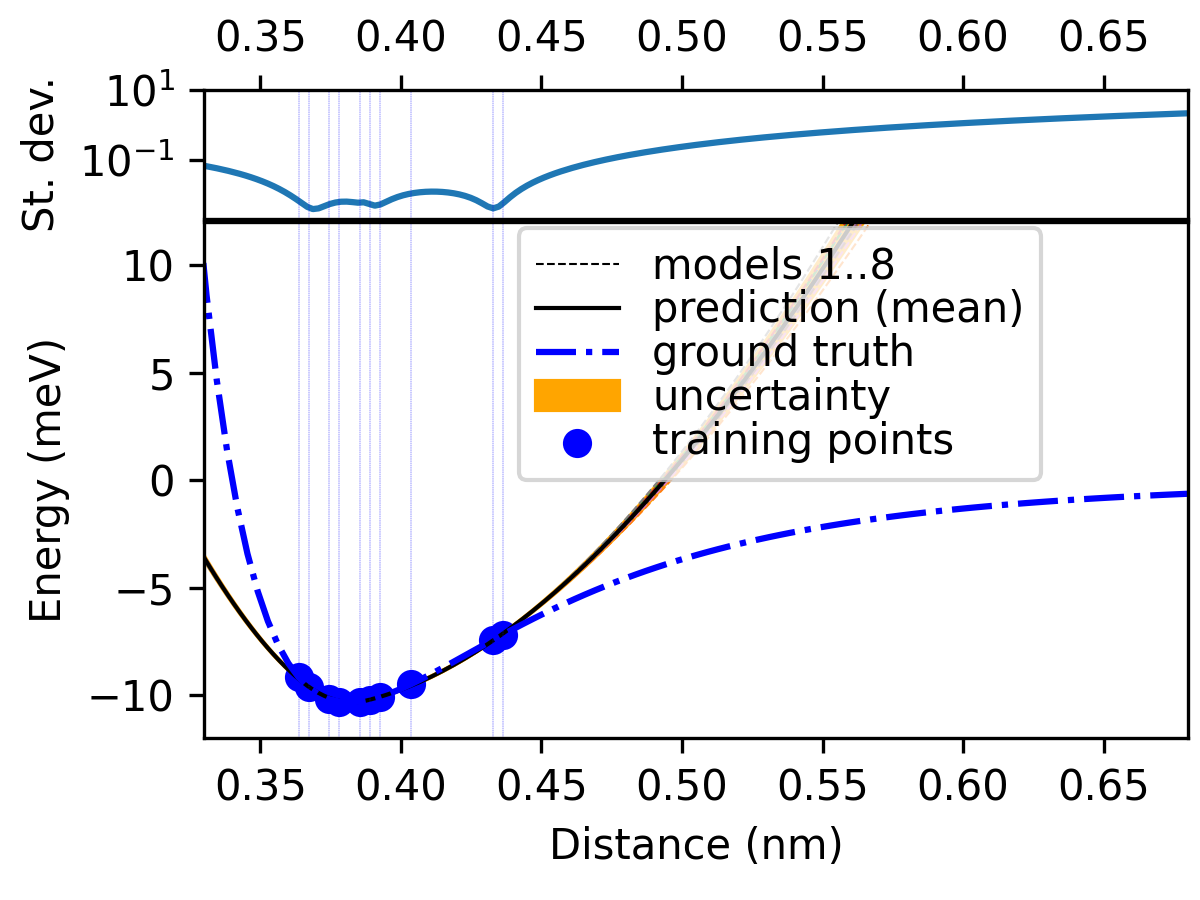}
    \includegraphics[width=0.49\hsize]{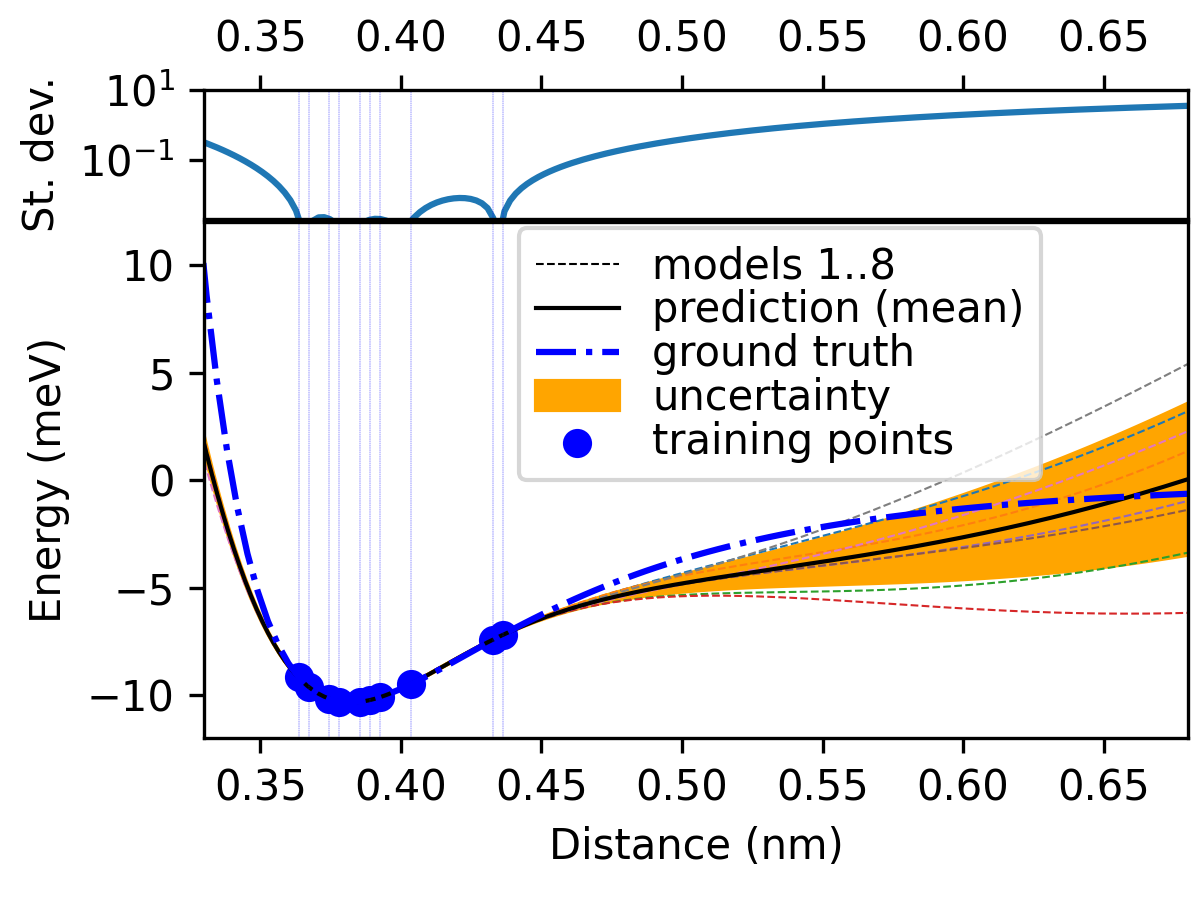}
    \caption{The ground truth (blue dash-dotted lines) and the training points (blue dots and dotted vertical lines) are shown together with prediction from an ensemble of eight models trained on the same data in the dimer case study.
    In the top left panel the NNPs  use the Sigmoid function  as their non-linear activation functions, and in the top right panel Rectified Linear Units (ReLU) are used.
    In the bottom left panel, the NNPs use the Continuously Differentiable Exponential Linear Units (CELU) as non-linearity.
    Gaussian Error Linear Units (GELU) are used in the bottom right panel.
    }
    \label{fig:lj-sigmoid-relu}
\end{figure}

\begin{figure}[ht]
    \centering
    \includegraphics[width=\hsize]{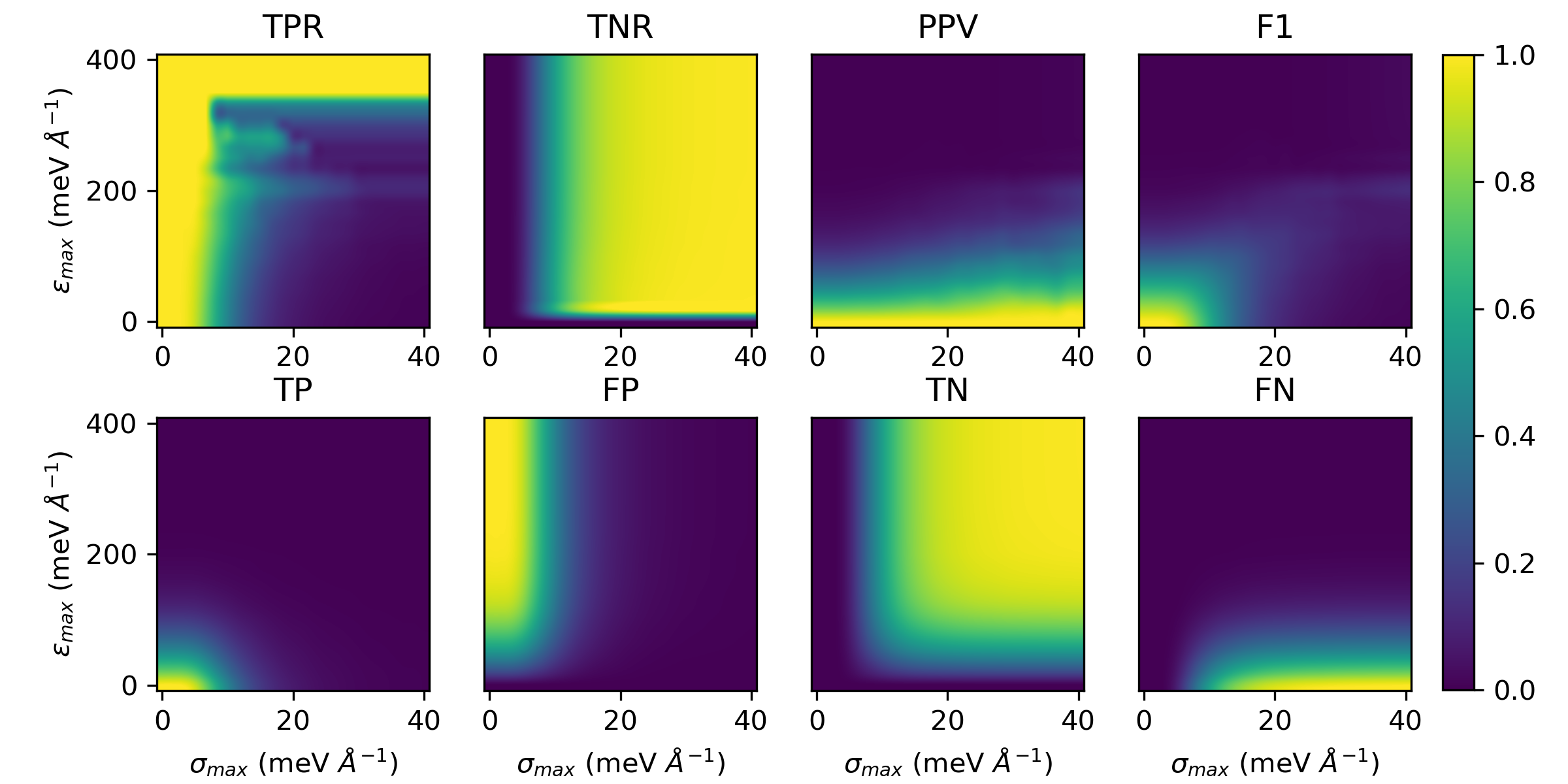}
    \caption{From left to right and top to bottom, we plot the
    recall/sensitivity/true positive rate (TPR),
    specificity/true negative rate (TNR),
    precision/positive predictive value (PPV),
    F$_1$ score (F1),
    true positive rate (TP),
    false positive rate (FP),
    true negative rate (TN),
    and false negative rate (FN)
    as heat maps  on the validation set $\mathcal{D}^{Al}_{1500}$.
    }
    \label{fig:heat-al}
\end{figure}

\begin{figure}[ht]
    \centering
    \includegraphics[width=\hsize]{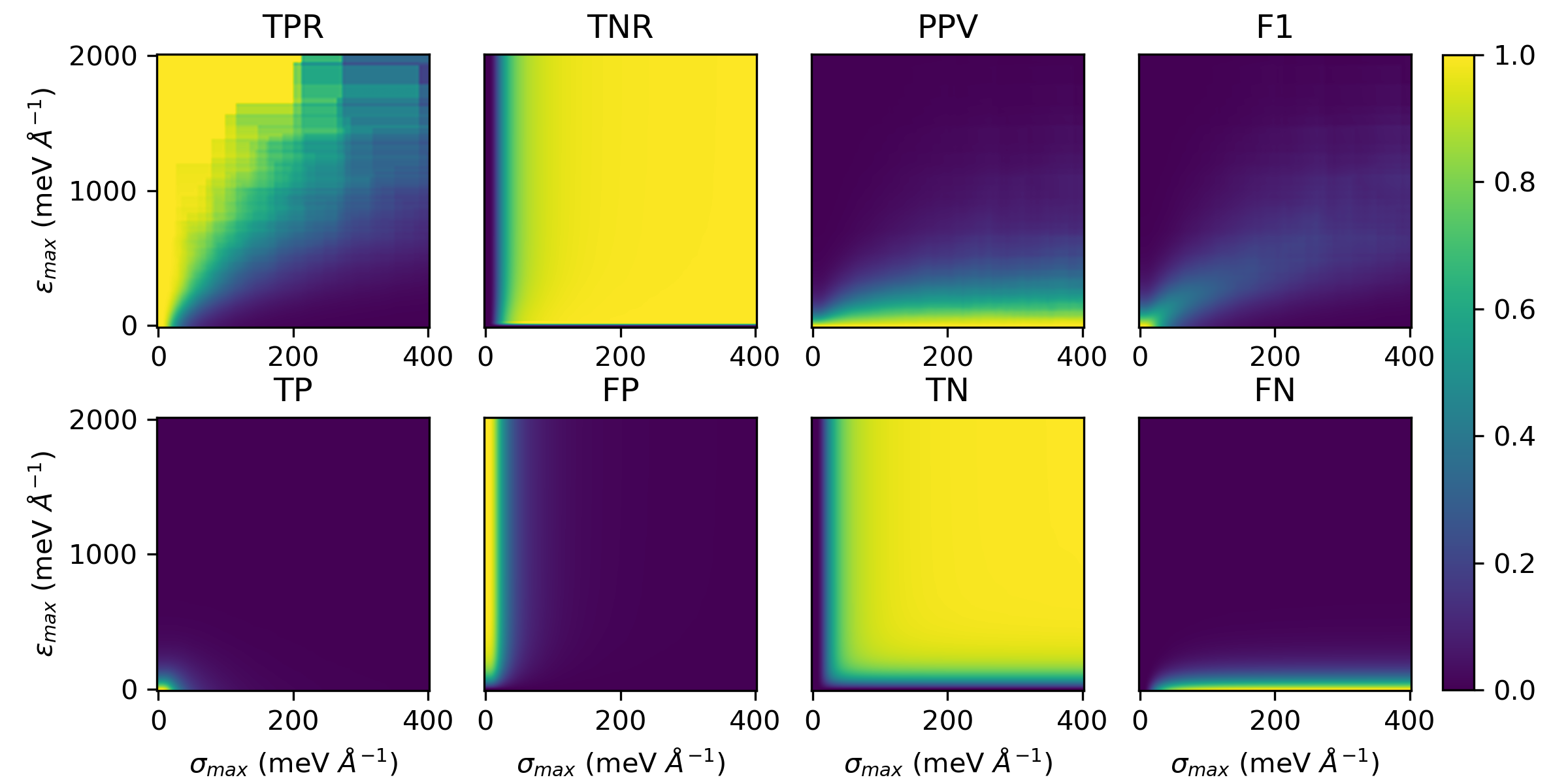}
    \caption{From left to right and top to bottom, we plot the
    recall/sensitivity/true positive rate (TPR),
    specificity/true negative rate (TNR),
    precision/positive predictive value (PPV),
    F$_1$ score (F1),
    true positive rate (TP),
    false positive rate (FP),
    true negative rate (TN),
    and false negative rate (FN)
    as heat maps  on the validation set $\mathcal{D}^{H_2O}_{3}$.
    }
    \label{fig:heat-water}
\end{figure}
\begin{figure}[ht]
    \centering
    \includegraphics[width=\hsize]{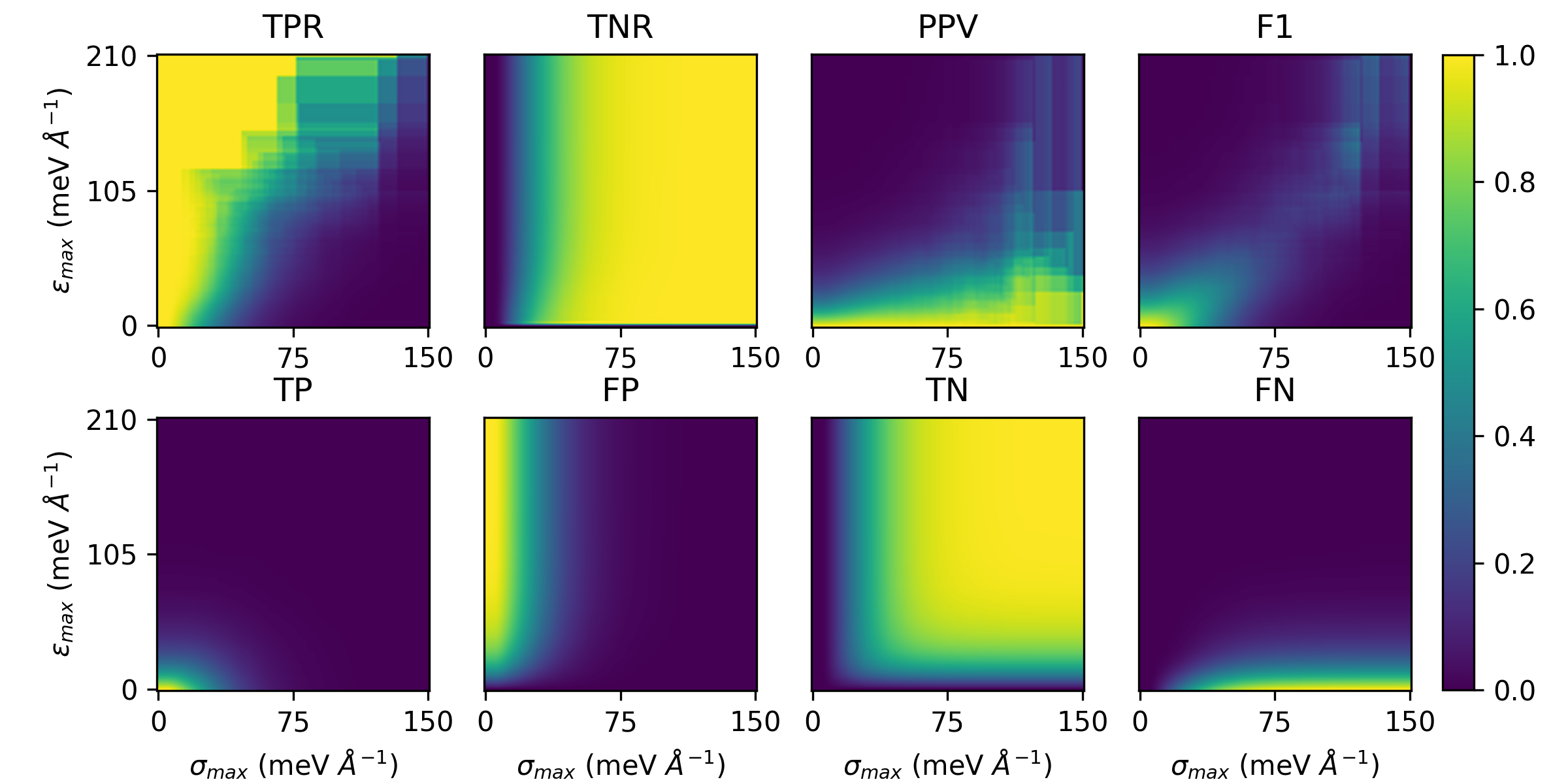}
    \caption{From left to right and top to bottom, we plot the
    recall/sensitivity/true positive rate (TPR),
    specificity/true negative rate (TNR),
    precision/positive predictive value (PPV),
    F$_1$ score (F1),
    true positive rate (TP),
    false positive rate (FP),
    true negative rate (TN),
    and false negative rate (FN)
    as heat maps  on the validation set $\mathcal{D}^{C_6H_6}_{5}$.
    }
    \label{fig:heat-benzene}
\end{figure}

%Figure \ref{fig:water_energy_distribution} illustrates the energy distribution of the four sets of configurations D$_0..3$. We used the data set provided by Cheng and coworkers consisting of configurations containing 64 water molecules. We sorted into ascending order these configurations using their potential energy. Then, we split the sorted configurations into four groups D$_0..3$. Each group contains the same number of structures. ]
 
\begin{figure}[ht]
    \centering
    \includegraphics[width=\hsize]{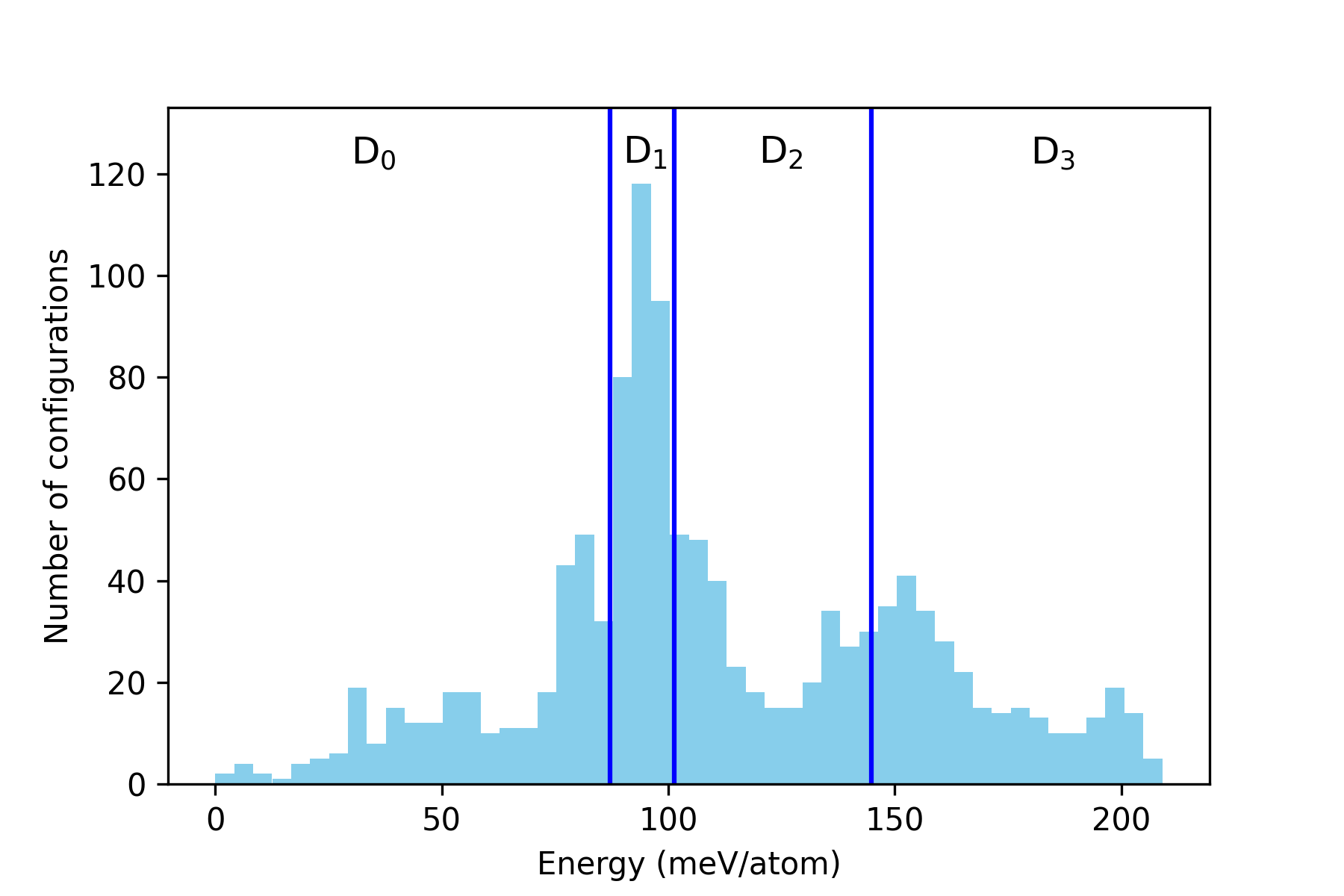}
    \caption{We plot the energy distributions of the four sets of bulk liquid H$_2$O configurations. 
    We sorted into ascending order these configurations using their potential energy and split in four sets: D$_0$, D$_1$, D$_2$, and D$_3$. Each set consists of 300 configurations of 64 water molecules. 
    %Vertical lines indicate the energy separating the sets.
    The potential energy values separating the four sets are indicated by three vertical lines at energies 87, 101, and 145~meV/atom. The zero is equal to the lowest energy of the set D$_0$.
    }
    \label{fig:water_energy_distribution}
\end{figure}

\begin{figure}[ht]
    \centering
    \includegraphics[width=\hsize]{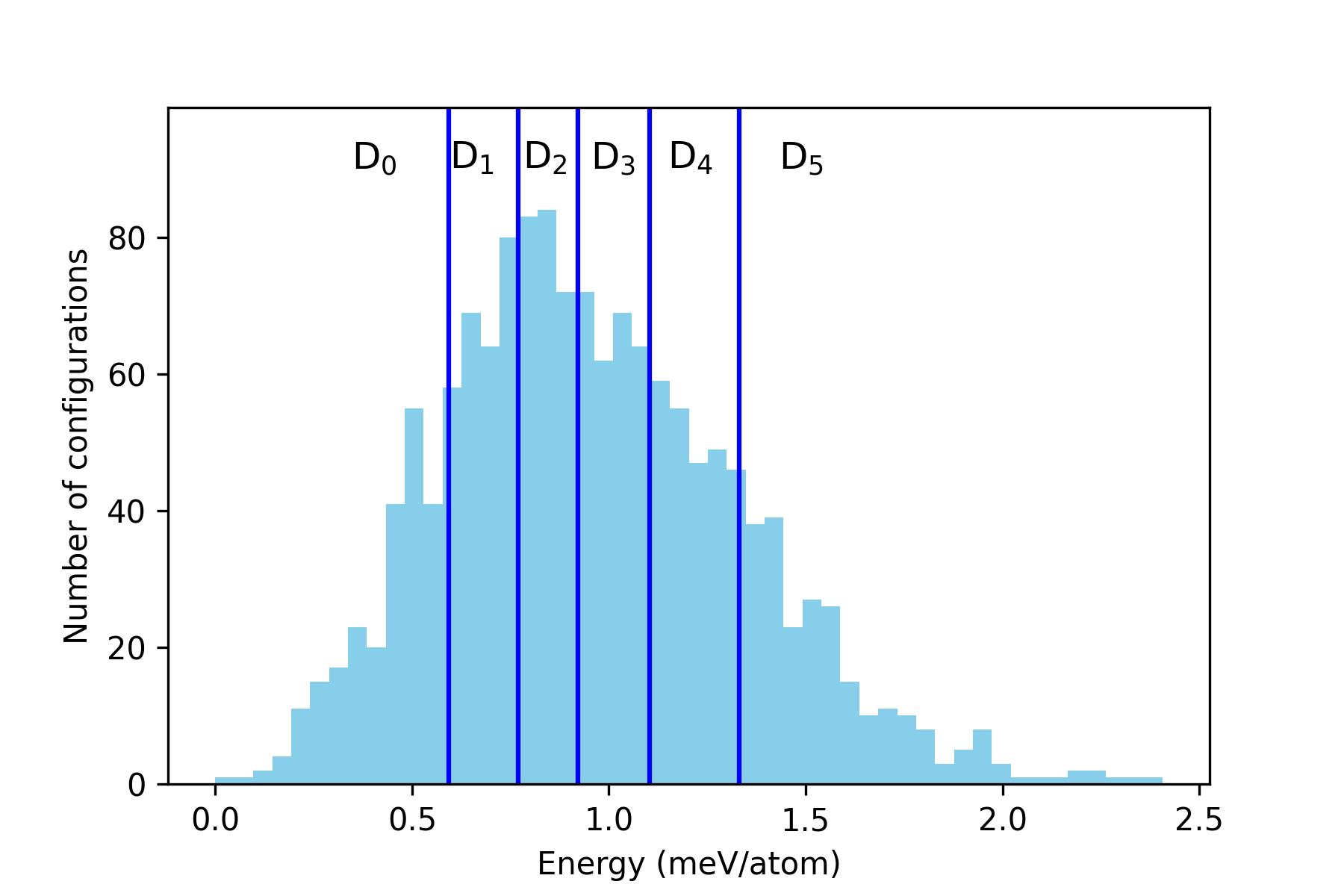}
    \caption{We plot the energy distributions of the six sets of C$_6$H$_6$ configurations. 
    We sorted into ascending order these configurations using their potential energy and split in six sets: D$_0..5$. Each set consists of 250 configurations. Vertical lines indicate the energy separating the sets.
    }
    \label{fig:benzene_energy_distribution}
\end{figure}

\begin{figure}
	\raisebox{20pt}{\includegraphics[width=0.42\hsize]{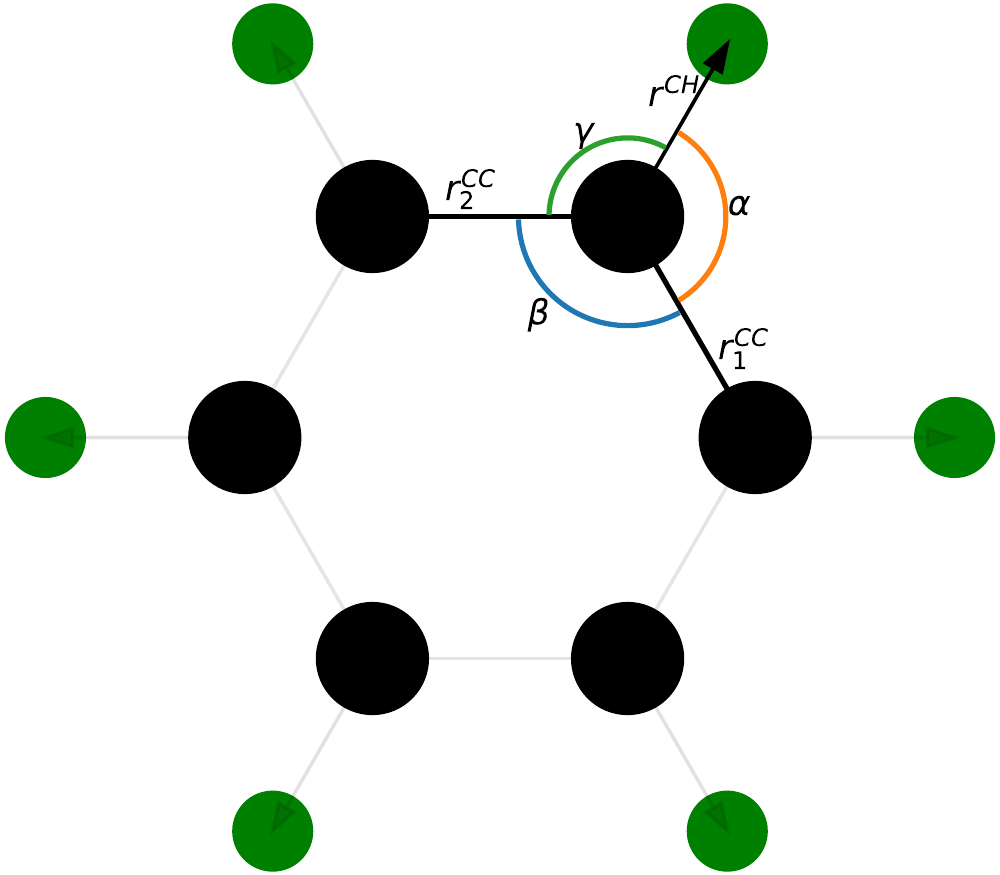}}	 
	\includegraphics[width=0.56\hsize]{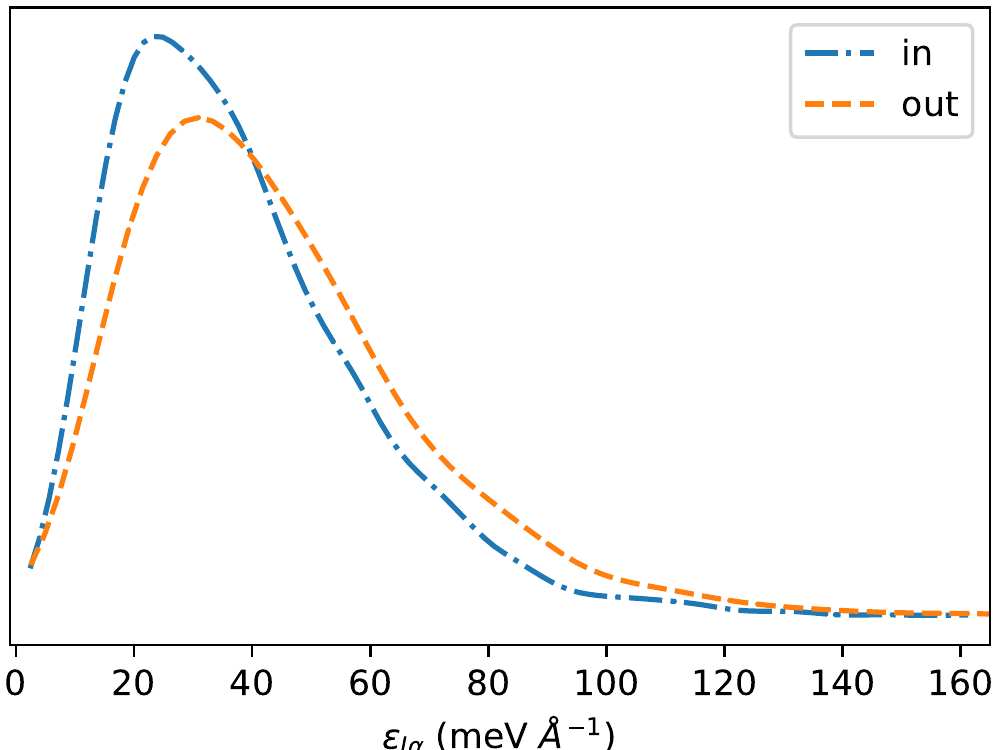}
\caption{
Left: schematic drawing of C$_6$H$_6$ with the C--C distances $r^{CC}_1$ and $r^{CC}_2$, the C--H distance $r^{CH}$ and the three angles $\alpha$, $\beta$, and $\gamma$ around one carbon atom.
For every carbon atom, we calculate a six-dimensional descriptor given by: 1) $r_{CH}$, 2) $0.5 (r^{CC}_1 + r^{CC}_2)$, 3) $(r^{CC}_1 - r^{CC}_2)^2$, 4) $\beta$, 5) $0.5(\alpha+\gamma)$, 6) $(\alpha-\gamma)^2$.
We take all the descriptors of all C--atoms in the training set $D_0^{C_6H_6}$ and calculate their convex hull in 6 dimensions.
For every descriptor of every carbon atom in the validation sets $D_{1\dots 3}^{C_6H_6}$, we find whether it lies inside or outside the convex hull given by the training data.\\
Right: We calculate the histogram of the true error given for every carbon atom whose descriptor has been found inside the convex hull (blue dash-dotted line) and outside (orange dashed line). Our findings suggest that, as expected, the error of the model increases when inferring the forces of configurations that cannot be interpolated well from training configurations.
}
\end{figure}

\end{document}